\documentclass[a4paper,11pt]{article}
\pdfoutput=1

\usepackage{graphicx}  
\usepackage{dcolumn}   
\usepackage{bm}        
\usepackage{amssymb}   
\usepackage{amsmath}   
\usepackage{multirow}
\usepackage{units}
\usepackage{xspace}
\usepackage[normalem]{ulem}
\usepackage{subcaption}

\usepackage{jinstpub}

\newcommand{\nova}{NOvA\xspace}
\newcommand{\genie}{GENIE\xspace}
\newcommand{\numi}{NuMI\xspace}
\newcommand{\numu}{\ensuremath{\nu_{\mu}}\xspace}
\newcommand{\nue}{\ensuremath{\nu_{e}}\xspace}

\newcommand{\nueCC}{\nue~{\rm CC}\xspace}
\newcommand{\numuCC}{\numu~{\rm CC}\xspace}
\newcommand{\nutau}{\ensuremath{\nu_{\tau}}\xspace}

\newcommand{\anue}{\ensuremath{\overline{\nu}_{e}}\xspace}          
\newcommand{\anumu}{\ensuremath{\overline{\nu}_{\mu}}\xspace}       

\newcommand{\nutauCC}{\nutau~{\rm CC}\xspace}

\newcommand{\xview}{$X$-view\xspace}
\newcommand{\yview}{$Y$-view\xspace}

\newcommand{\relu}{ReLU\xspace}
\newcommand{\googlenet}{GoogLeNet\xspace}

\title{A Convolutional Neural Network Neutrino Event Classifier}

\author[a,1]{A. Aurisano,%
\note{Corresponding authors.}}
\author[b,1]{A. Radovic,}
\author[c,1]{D. Rocco,}
\author[d]{A. Himmel,}
\author[e]{M.D. Messier,}
\author[d]{E. Niner,}
\author[c]{G. Pawloski,}
\author[e]{F. Psihas,}
\author[a]{A. Sousa}
\author[b]{ and P. Vahle}

\affiliation[a]{University of Cincinnati,\\
  Cincinnati, Ohio 45221, USA}

\affiliation[b]{College of William \& Mary,\\
  Williamsburg, Virginia 23187, USA}

\affiliation[c]{University of Minnesota,\\
  Minneapolis, Minnesota 55455, USA}

\affiliation[d]{Fermi National Accelerator Laboratory,\\
  Batavia, Illinois 60510, USA}

\affiliation[e]{Indiana University,\\
  Bloomington, Indiana 47405, USA}

\emailAdd{aurisaam@ucmail.uc.edu, aradovic@wm.edu, rocco@physics.umn.edu}

\abstract{ Convolutional neural networks (CNNs) have been widely
  applied in the computer vision community to solve complex problems
  in image recognition and analysis. We describe an application of the
  CNN technology to the problem of identifying particle interactions in sampling
  calorimeters used commonly in high energy physics and high energy
  neutrino physics in particular. Following a discussion of the core
  concepts of CNNs and recent innovations in CNN architectures related
  to the field of deep learning, we outline a specific application to
  the NOvA neutrino detector. This algorithm, CVN (Convolutional
   Visual  Network) identifies neutrino
  interactions based on their topology without the need for detailed 
  reconstruction and outperforms algorithms currently in use by the
  NOvA collaboration.
}


\begin{document}
\maketitle
\flushbottom



\section{Introduction} \label{sec:intro}

A core problem in experimental high-energy particle physics (HEP) is
the correct categorization of the particle interactions recorded in
our detectors as signal and background. Commonly, this
characterization has been done by reconstructing high-level components
such as clusters, tracks, showers, jets, and rings associated with
particle interactions recorded by the detector and summarizing the
energies, directions, and shapes of these objects with a handful of
quantities. These quantities are then either directly selected or fed
into machine learning algorithms such as K-Nearest
Neighbors~\cite{knn}, Boosted Decision
Trees~\cite{friedman2002stochastic}, or Multilayer
Perceptrons~\cite{Rosenblatt1961, reed1999neural} to separate signal
from background.  While these techniques have been very successful,
they are prone to two potential failings: mistakes in the
reconstruction of high level features from the raw data can lead to
incorrect categorization of the physics event, and the features used
to characterize the events are limited to those which have already
been imagined and implemented for the experiment.

This core problem shares many similarities with the problems
confronted in computer vision. As in HEP, the computer vision
community has explored many approaches to extract specific features
from images to enable categorization. Recently, however, computer
vision has made great advances by moving away from using specifically
constructed features to the extraction of features using a machine
learning algorithm known as a convolutional neural network
(CNN)~\cite{LeCun1989}.

CNNs are well suited to a broad class of detectors used in HEP and
particularly in high energy neutrino physics. Sampling calorimeters
that use scintillator (e.g. NOvA~\cite{nova}, MINERvA~\cite{minerva}
and MINOS~\cite{minos}), liquid argon time projection chambers
(e.g. ICARUS~\cite{Bettini:1991fh}, MicroBooNE~\cite{microboone},
DUNE~\cite{dune1,dune2,dune3,dune4}), and water Cherenkov detectors
(e.g. IceCube~\cite{icecube1, icecube2} and
Super-Kamiokande~\cite{superk}) record the amount of energy deposited
in small regions throughout the volume of the detector. When these
measurements are combined, they result in what is essentially an image
of the physics interaction which is well suited to analysis using
computer vision tools. Early studies with Daya Bay data~\cite{Guo:2007ug, Racah:2016gnm} and
simulated LHC jets~\cite{Aad:2008zzm, Chatrchyan:2008aa, deOliveira:2015xxd}
have demonstrated that CNNs can be powerful tools in high energy physics.

In this paper, we describe CNNs and the techniques commonly used to
build and train these networks. We then outline the construction,
training, and testing of a specific application of the CNN technique,
``CVN'' (Convolutional Visual Network), to events simulated in the
\nova experiment~\cite{nova} and present benchmarks of its
performance.

\section{Deep Learning and Convolutional Neural Networks} \label{sec:deepLearning}

The multilayer perceptron (MLP)~\cite{Rosenblatt1961, reed1999neural},
or traditional neutral network, is a machine learning algorithm in
wide use in HEP. The structure of an MLP consists of an input layer,
one or more hidden layers, and an output layer.  The goal of an MLP is
to approximate a function $f: \mathbb{R}^{n} \to \mathbb{R}^{m}$,
where $n$ is the dimensionality of the input $\vec{x}$ and $m$ is the
dimensionality of the output $\vec{f}$. All layers in traditional MLPs
are fully connected, meaning that the output of each node is the
weighted sum of the outputs of all nodes in the previous layer plus a
bias term, operated on by a non-linear function. Traditionally, the
preferred non-linearity is a sigmoid function such as \textit{tanh} or
the logistic function~\cite{reed1999neural}. An MLP with a single
hidden layer, under certain assumptions, can be shown to approximate
any function to arbitrary precision given a sufficient number of
hidden nodes~\cite{Hornik, Cybenko}. The weights and biases used in an
MLP are typically determined using supervised learning. During
supervised learning~\cite{Rumelhart:1986:PDP:104279}, the MLP is
presented examples where both $\vec{x}$ and the corresponding output,
$\vec{f}$, referred to as the \textit{ground truth}, are known.  The
\textit{loss}, a measure of the error between the output of the MLP
and the ground truth is computed, and its gradient as a function the
weights and biases is calculated using the \textit{back-propagation}
algorithm~\cite{hinton1986}.  The loss is then minimized by altering
the weights and biases using the stochastic gradient
descent~\cite{LeCun1998} method. This procedure is repeated until the
errors are reduced to an acceptable level.

The MLP is a powerful technique, but it has a number of
deficiencies~\cite{Bengio-2009}. First, it tends to scale poorly to a
large number of raw inputs. Historically, most of the work in
developing an MLP for a particular task was devoted to extracting
features from the raw data that could be used as optimal
inputs~\cite{lecun2015deep}. In HEP, this is essentially the process
of reconstruction; however, developing optimal, robust reconstruction
routines is difficult and time consuming.  Second, although a single
hidden layer can approximate most functions to arbitrary precision,
the number of nodes necessary in that hidden layer may approach
infinity. Networks with multiple hidden layers can often reach the
required accuracy with fewer nodes than the equivalent single layer
network~\cite{reed1999neural}. However, multilayer networks can be
difficult to train. This is partially due to the fact that sigmoid
functions are saturating, that is, as the input to the sigmoid
approaches $\pm\infty$, the gradient approaches zero. The updates to
the weights and biases applied using the stochastic gradient descent
method have a term proportional to the gradient, so this situation can
slow down or halt learning.  In shallow networks, this can be
controlled through the careful preparation of inputs~\cite{LeCun1998},
but it is difficult to keep the inputs to nodes in the non-saturated
range over many layers.  Third, the large number of free parameters in
a large network runs the risk of over-training in which the network
learns to reproduce the training sample too well and fails to
generalize to inputs it has not seen~\cite{reed1999neural}.

Deep learning~\cite{lecun2015deep}, the use of architectures with many
layers, has had considerable success in tasks like image
recognition~\cite{krizhevsky2012imagenet, szegedy2014googlenet} and
natural language processing~\cite{farabet-pami-13} and has been made
possible by several advances that mitigate the deficiencies of
traditional MLPs.  Instead of relying on engineered features as
inputs, the development of structures like CNNs have made it possible
to robustly and automatically extract learned features.  To allow for
the efficient training of deep structures, saturating non-linearities
are frequently replaced by rectified linear
units (\relu)~\cite{icml2010_NairH10}, defined as $f(x) = max(0,x)$,
which is non-saturating. Finally, over-training is mitigated in fully
connected layers using the regularization technique called
\textit{dropout}~\cite{hinton2014dropout} in which, at every training
iteration, each weight is set to zero with a probability $r$ while the
remaining weights are scaled up by a factor of $1/(1-r)$ to roughly
maintain the overall scale of the value passed through the
non-linearity. In this way, every iteration only uses a random
subsample of the possible connections, leading to a final network
which is effectively an ensemble of smaller networks.

In this paper we will focus on CNNs, which have been highly successful
in the field of computer vision for classification and other
tasks~\cite{lecun2010convolutional,krizhevsky2012imagenet}. The
technique was inspired by studies of the visual cortex of
animals~\cite{hubel68}. In these studies, it was found that the visual
cortex contains \textit{simple cells}, which are sensitive to
edge-like features within small regions of the retina, and
\textit{complex cells}, which are receptive to collections of simple
cells and are sensitive to position independent edge-like
features. These structures can be modeled by performing discrete
convolutions to extract simple features across the visual field.  CNNs
mimic this structure using a series of convolutional layers that
extract a set of features from the input image and pooling layers that
perform dimensionality reduction and add translational invariance.

The data passed from layer to layer in a CNN has a three dimensional
structure - height, width, and channel number. Height and width refer
to the dimensions of the input image, and channel number is defined in
analogy with the RGB channels of color images. For an $n \times m$
convolutional layer, the input data is transformed according to,
  \begin{equation}
    (f*g)_{p,q,r} = \sum_{i = 1}^{n}\sum_{j = 1}^{m}\sum_{k = 1}^{c} f_{i,j,k,r} g_{p+i,q+j,k},
  \end{equation}
where $(f*g)_{p,q,r}$ refers to the $(p,q)$ pixel of the $r$ channel of the transformed image, $n$ and $m$ are the height and width of the convolutional kernel, $c$ is the number of channels of the input image, $f$ is a filter, and $g$ is an array corresponding to pixel intensities of the input image. The filter $f$ is a four dimensional tensor where $i$ and $j$ index the height and width of the filter, $k$ indexes the input channel, and $r$ indexes the output channel, and it is trained to identify features within the image. For a fixed $k$ and $r$, the filter, $f$, can be thought of as an $n \times m$ convolutional kernel.  After applying a separate convolutional kernel to each channel and performing a weighted sum across channel, the resulting output image is known as a \textit{feature map}. The range of the $r$ dimension determines the number of $c$ stacks of $n \times m$ convolutional kernels that are trained. Each of these stacks of kernels produces a feature map which are stored in the channel dimension of the layer output. Finally, each output pixel is operated on by a non-linear function.

In this way,
convolutional layers~\cite{lecun2015deep} produce many alternative
representations of the input image, each serving to extract some
feature which is learned from the training sample. At early stages, 
the feature maps often resemble the original image with certain elements
emphasized, but they become more abstract at later stages of the network.

Since each convolutional layer generates many feature maps which have
comparable dimensions to the original input image, the memory
requirements and number of operations needed to evaluate the network
can grow dramatically. Pooling is a technique to down-sample the size
of feature maps; we have made use of two pooling techniques
\textit{max pooling} and \textit{average
  pooling}~\cite{lecun2010convolutional}. In $n \times m$ \textit{max
  pooling}, the image is down-sampled by replacing an $n \times m$
region of the image with a single value corresponding to the maximum
value in that region; in \textit{average pooling} the average value is
used. The pooled regions may be chosen to
overlap~\cite{krizhevsky2012imagenet,szegedy2014googlenet} to reduce
information loss. Since each pixel after pooling corresponds to $n
\times m$ before pooling, small translations in input features result
in identical output. This decreases the network's sensitivity to the
absolute location of elements in the image.

\begin{figure}[tbp]
\begin{center}
\includegraphics[width=0.8\textwidth]{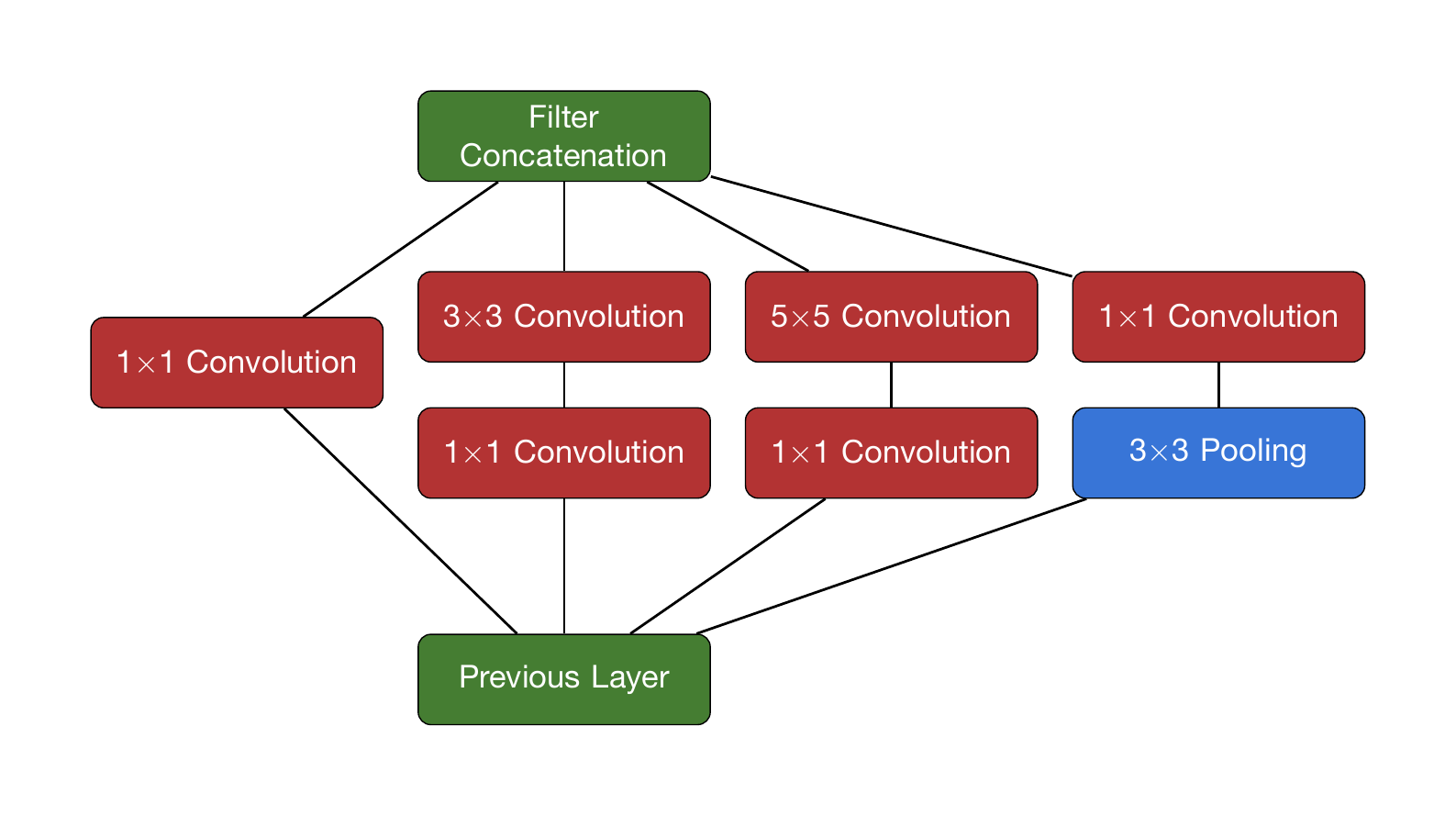}
\end{center}
\caption{Diagram of the inception module}{Like any single layer, the
  inception module takes the set of feature maps produced by the
  previous layer as input. It then distributes those feature maps to 
  branches, each with filters at different scales.  NIN architecture is
  implemented with $1\times1$ convolutions  to down-sample into a
  smaller number of maps, maintaining the dimensionality of the input maps. 
  Separate branches perform $3\times3$
  and $5\times5$ convolution, as well as $3\times3$ overlapping
  pooling. The filtered outputs from each branch are then concatenated to
  produce an output to the next layer with the same number of feature 
  maps, each with the same dimensions, as were passed as input to the inception module.
}
\label{inception}
\end{figure}

The network we will describe in this paper is inspired by the
\googlenet~\cite{szegedy2014googlenet} architecture, which excelled at
the ImageNet image classification task \cite{ILSVRC15}. The core of
\googlenet's power comes from its use of the network-in-network
(NIN)~\cite{lin2013network} approach to augment the learning capacity
of convolutional layers while also reducing dimensionality. In NIN the
main network is composed of repeating sub-networks, where each
sub-network resembles a complete conventional CNN with convolution
layers at a variety of scales to capture complex behavior. To avoid
exponentially increasing the number of feature maps, NINs use a
convolutional layer applying $1\times1$ convolutional kernels. This
performs a weighted sum over feature maps to down-sample into a
smaller number of maps. The sub-network in the \googlenet
architecture, called the \textit{inception module}, is shown in
Figure~\ref{inception}. Each branch of the inception module applies
filters which extract features of various scales.  The \googlenet
architecture also makes use of the technique \textit{local response
  normalization} (LRN) in which the response of a given cell in a
kernel map is normalized relative to the activity of adjacent kernel
maps. This creates competition for large valued features between
outputs computed by different kernels which helps the network avoid
local minima and to converge to a more optimal set of weights.

\section{Application to \nova event classification} \label{sec:appToNova}

\begin{figure}[htb]
\begin{center}
\includegraphics[width=0.8\textwidth]{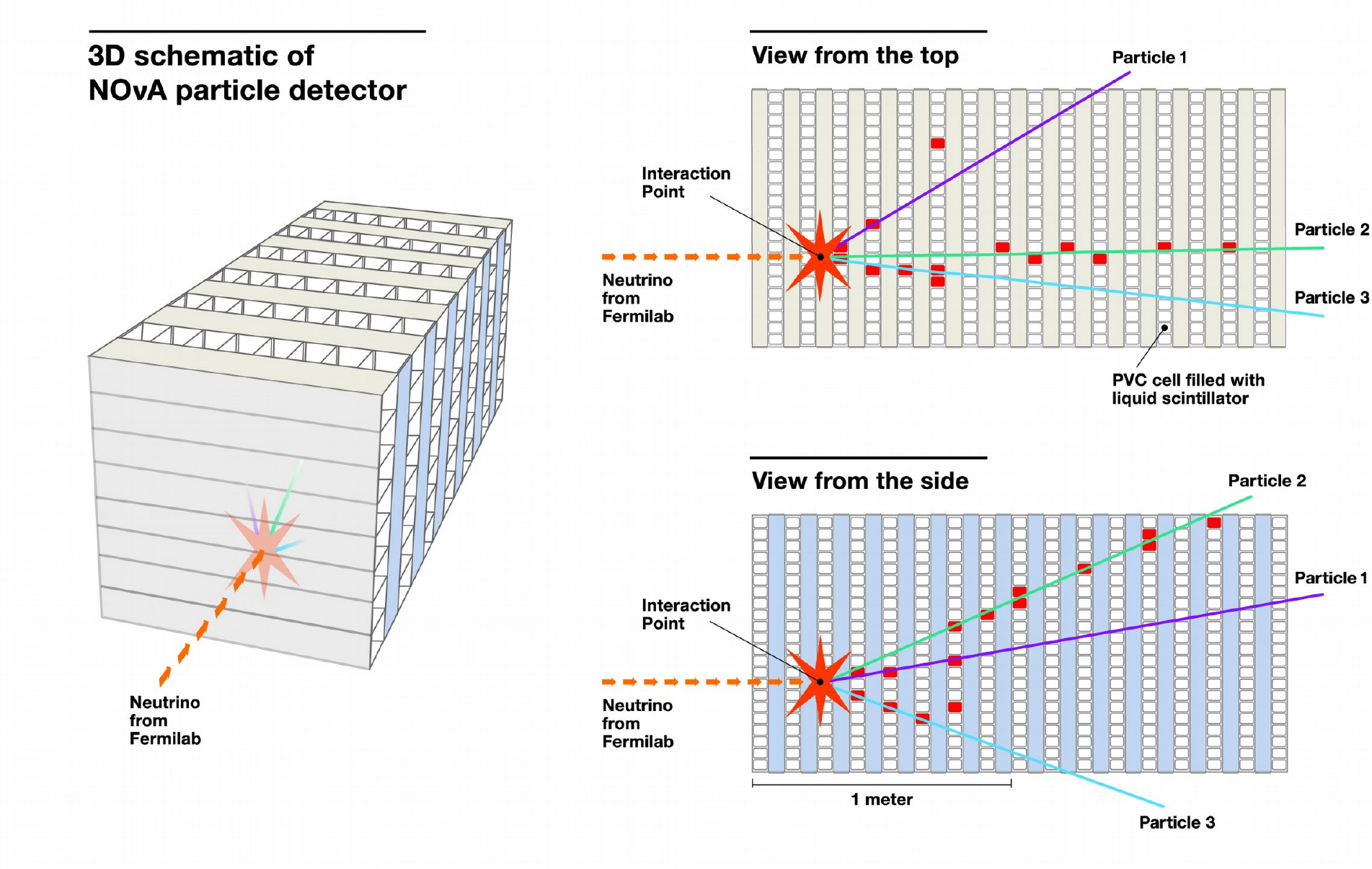}
\end{center}
\caption{\label{schematic} Schematic of the \nova~detector design}{The
  two figures on the right show the views through the top and side of
  the three-dimensional figure on the left. They show the `hits'
  produced as charged particles pass through and deposit energy in the
  scintillator-filled cells. Illustration courtesy of Fermilab.}
\end{figure}

We have developed and trained our own CNN, ``CVN'', for the
identification of neutrino events recorded by the NOvA
experiment. \nova aims to make precision measurements of neutrino
oscillation parameters via the disappearance of \numu and appearance
of \nue from neutrino oscillation.  \nova consists of two functionally
identical detectors in the \numi (Neutrinos at the Main Injector)
beam~\cite{numi} at Fermilab which produces a focused beam with an
initial flavor composition largely dominated by \numu and a small
intrinsic \anumu, \nue, and \anue components.  Placing the detectors
off-axis at 14.6~mrad provides a narrow-band neutrino energy spectrum
near 2~GeV.  The Near Detector, located at Fermilab, is placed 1~km
from the neutrino source; the Far Detector is located 810~km away near
Ash River, Minnesota.  The \nova detectors are composed of extruded
PVC cells filled with liquid scintillator which segment the detector
into cells with a cross section 3.9~cm wide $\times$ 6.6~cm deep.  The
cells are 15.5~m long in the Far Detector.
Scintillation light from charged particles can be captured by a
wavelength shifting fiber which runs through each cell.  The end of
the fiber is exposed to a single pixel on an avalanche photo-diode
array to record the intensity and arrival time of photon signals.  The
spatial and absolute response of the detector to deposited light is
calibrated out using physical standard candles, such that a calibrated
response can be derived which is a good estimate of the true deposited
energy.  Parallel cells are arrayed into planes, which are configured
in alternating horizontal and vertical alignments to provide separate,
interleaved $X$-$Z$, and $Y$-$Z$ views. The 14,000~ton Far Detector,
which is used for the training and evaluation of CVN in this paper,
consists of 344,064 total channels arranged into 896 planes each 384
cells wide~\cite{nova}. Information from the two views can be merged
to allow 3D event reconstruction. A schematic of the detector design
can be seen in Figure~\ref{schematic}.

Reconstruction of the neutrino energy and flavor state at the detector
is essential to neutrino oscillation measurements.  The neutrino
flavor state can be determined in charged-current (CC) interactions
which leave a charged lepton in the final state; an electron in the
case of \nue, a muon in the case of \numu, or a tau in the case of
\nutau.  Neutral-current (NC) interactions bear no signature of
the flavor of the interacting neutrino and are thus a background for the
charged-current analyses, but may be signal events in other
searches. 

To support these analyses, we constructed the CVN identifier to
characterize candidate neutrino events into one of the following
interaction types.

\begin{itemize}
  \item \numuCC - A muon plus a hadronic component. One of the
    main topological features of these events is the long, low $dE/dx$
    track corresponding to the track of a minimally ionizing
    muon. 
  \item \nueCC - An electron plus a hadronic component. The
    electron topology is typically a wide shower, rather than a track,
    whose dimensions are related to the radiation length of the
    detector material. 
  \item \nutauCC - A tau plus a hadronic component.  The tau is extremely short lived and not visible in the 
   \nova detector but decays immediately with varying final state probabilities that may produce pions, electrons, muons, and 
    neutrinos.  The production threshold for these events is 3.4 GeV, at the upper end of the energy spectrum seen in the \nova 
    detectors.
  \item $\nu$ NC- The outgoing lepton in these interactions is a
    neutrino, which will travel onward undetected. Thus, only the
    hadronic component of these events is visible, making their flavor
    impossible to identify. 
\end{itemize}

While it is useful to think about each category as a particular iconic
topology, misidentification can still occur. 
In particular NC interactions can be mistaken for CC interactions when they 
produce pions which look like leptonic activity. 
A charged pion track can appear quite similar to a muon track, with the exception
of a spike in energy deposition at the end of the track.
A neutral pion will rapidly decay to produce a pair of photons which themselves produce electromagnetic
showers, which are difficult to distinguish from showers produced by an electron, unless you
can find the telltale gap between the interaction vertex and the shower. 
By constructing an identification algorithm like CVN, which views the entire event topology,
we hope to minimize these misidentification failure modes but they remain a challenge.

\begin{figure}
    \begin{center}
    \begin{subfigure}[c]{0.8\textwidth}
        \includegraphics[width=\textwidth]{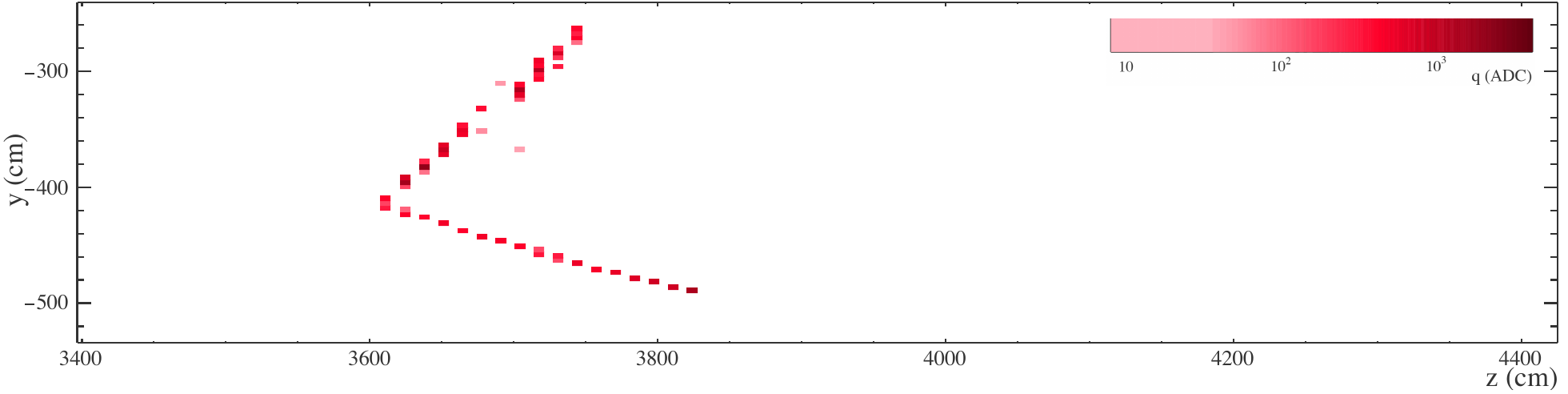}
        \caption{A \nueCC QE electron plus one hadron
          signature where the upward-going shower-like prong with multiple
          hit cells on each plane corresponds to an electron and
          the downward-going track-like prong with approximately one hit per plane
          correspond to a proton.\\ }
        \label{evd01}
    \end{subfigure}
    \begin{subfigure}[c]{0.8\textwidth}
        \includegraphics[width=\textwidth]{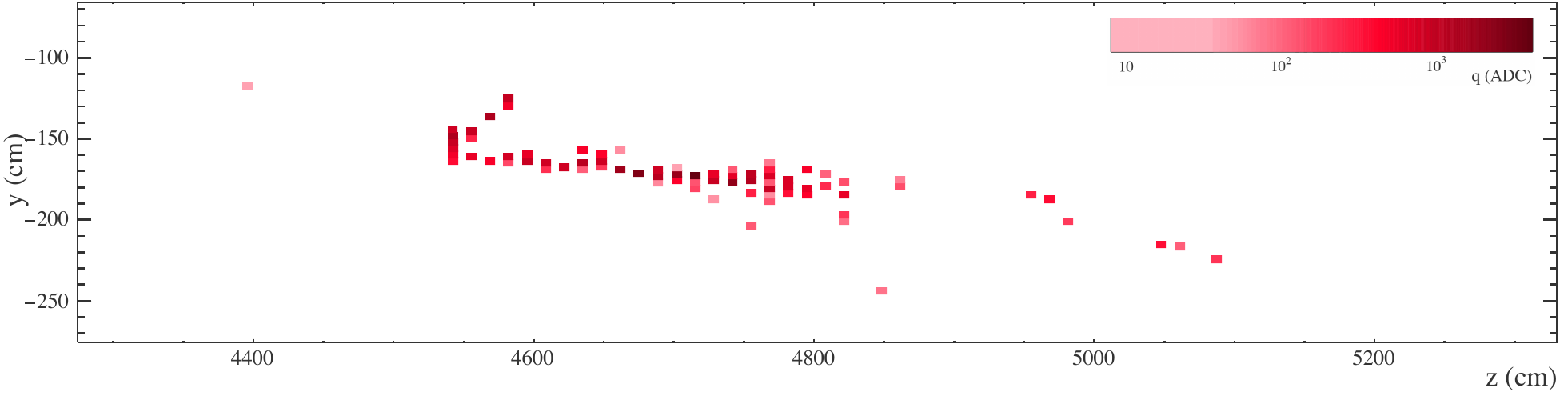}
        \caption{A \nueCC RES electron plus hadron shower
          signature with a characteristic electron shower and
          short prongs which could correspond to multiple hadrons.\\ }
        \label{evd02}
    \end{subfigure}
    \begin{subfigure}[c]{0.8\textwidth}
        \includegraphics[width=\textwidth]{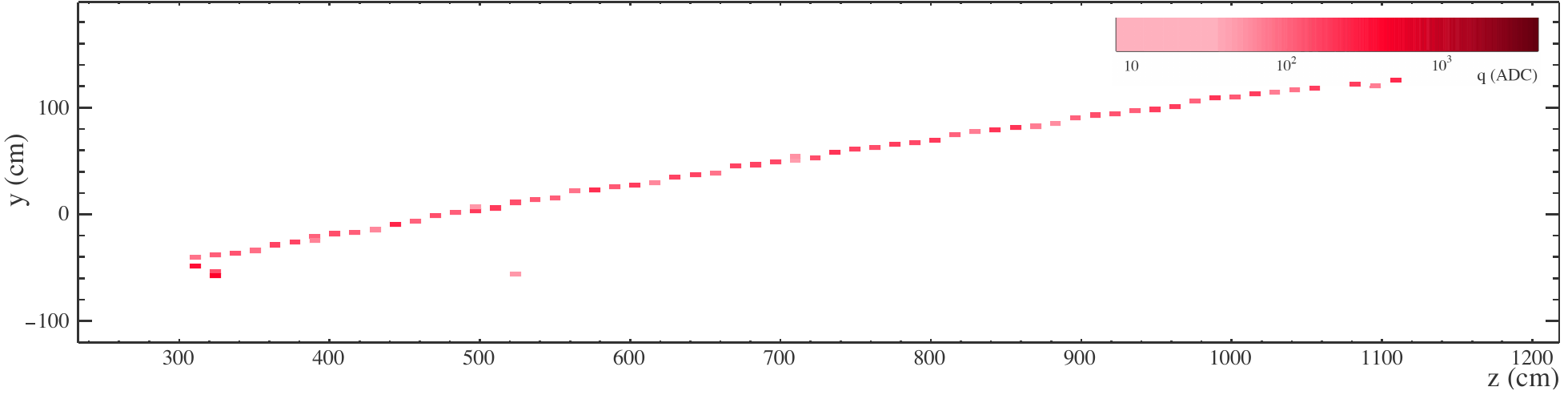}
        \caption{ A \numuCC QE muon plus one hadron
          signature with a long track-like prong with lower charger-per-cell
          corresponding to a muon and a short prong with larger charge-per-cell
          corresponding to a proton.\\ }
        \label{evd03}
    \end{subfigure}
    \end{center}
    \caption{Simulated events in \nova}{Each panel is the side view of a
      section of the \nova detector, as depicted in Figure
      \protect\ref{schematic}. The charge
      measured in each cell in units of ADC is indicated by color.}
      \label{nuevents}
\end{figure}

CC interactions were further divided into quasi-elastic (QE), resonant
(RES), and deep-inelastic-scattering (DIS) categories which vary in
the complexity of the hadronic portion of the event.  QE events are
two-bodied with the nucleon recoiling intact from the scattering
lepton.  In RES events the nucleon is knocked into a baryonic
resonance and decays back down to the nucleon with associated hadrons,
and in higher energy DIS events the nucleon breaks up in the process
of hadronization.  Figure~\ref{nuevents} shows example, simulated,
events from these categories as they might be recorded by the \nova
detectors. While the network shows some promise in being able to categorize
events at this detailed level, for now we have focused only on neutrino
event flavor identification by combining the outputs
of these detailed subdivisions into the four categories in the list above.

The input to CVN was formed first by clustering energy deposits recorded in each scintillator column together in space and time into \textit{slices}.  These slices efficiently separate deposits due to neutrino interactions from those due to cosmic ray interactions and remove nearly all energy deposits due to noise. Grids 100 planes deep and 80 cells wide are chosen which contain the slice. Two separate grids were made for the $x-z$ and $y-z$ detector
views. These grids are windows of the detector activity 14.52~m deep
and 4.18~m wide. The upstream side of
the window was chosen to align with the first plane that contains a
detector hit and the window is centered on the median hit cell
position. The size and placement of this window ensured that the
majority of neutrino events, including muon neutrino CC interactions,
are fully contained. The intensity of each pixel in this grid is
proportional to the calibrated energy deposition in each scintillator
column allowing these projections to naturally be interpreted as
grayscale images.

\begin{figure}[htbp]
  \begin{center}
    \includegraphics[width=0.6\textwidth]{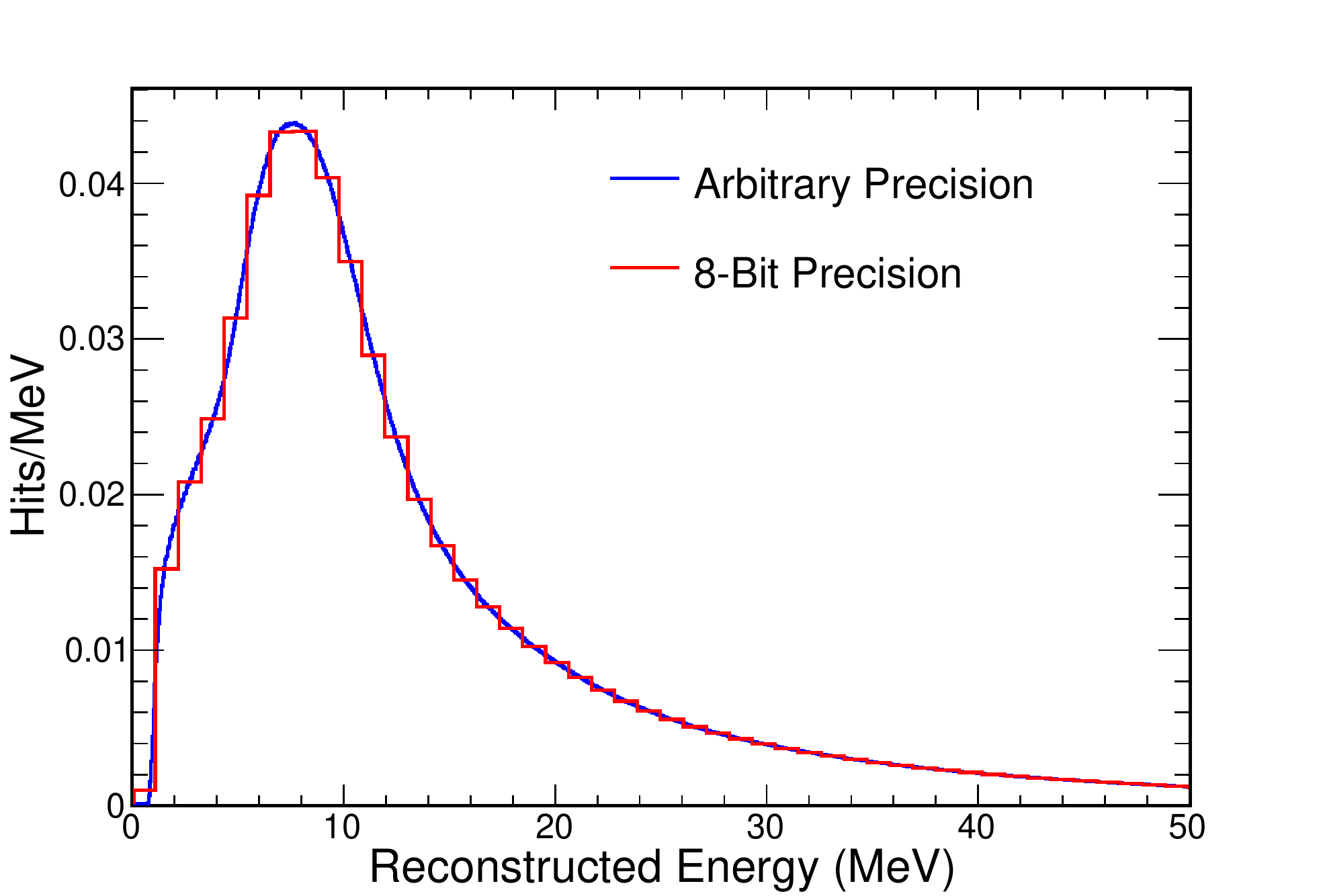}
    \caption{\label{fig:CompressedEnergy}A comparison of the energy spectrum of individual hits using arbitrarily fine binning and using 256 bins. This shows that it is acceptable to encode pixel intensities using 8-bits. In this encoding, the maximum representable energy per pixel is 278~MeV.}
    \end{center}
\end{figure}

\begin{figure*}[htbp]
  \begin{center}
    \begin{subfigure}[c]{\textwidth}
      \begin{center}
        \begin{subfigure}[c]{0.49\textwidth}
          \includegraphics[width=\textwidth]{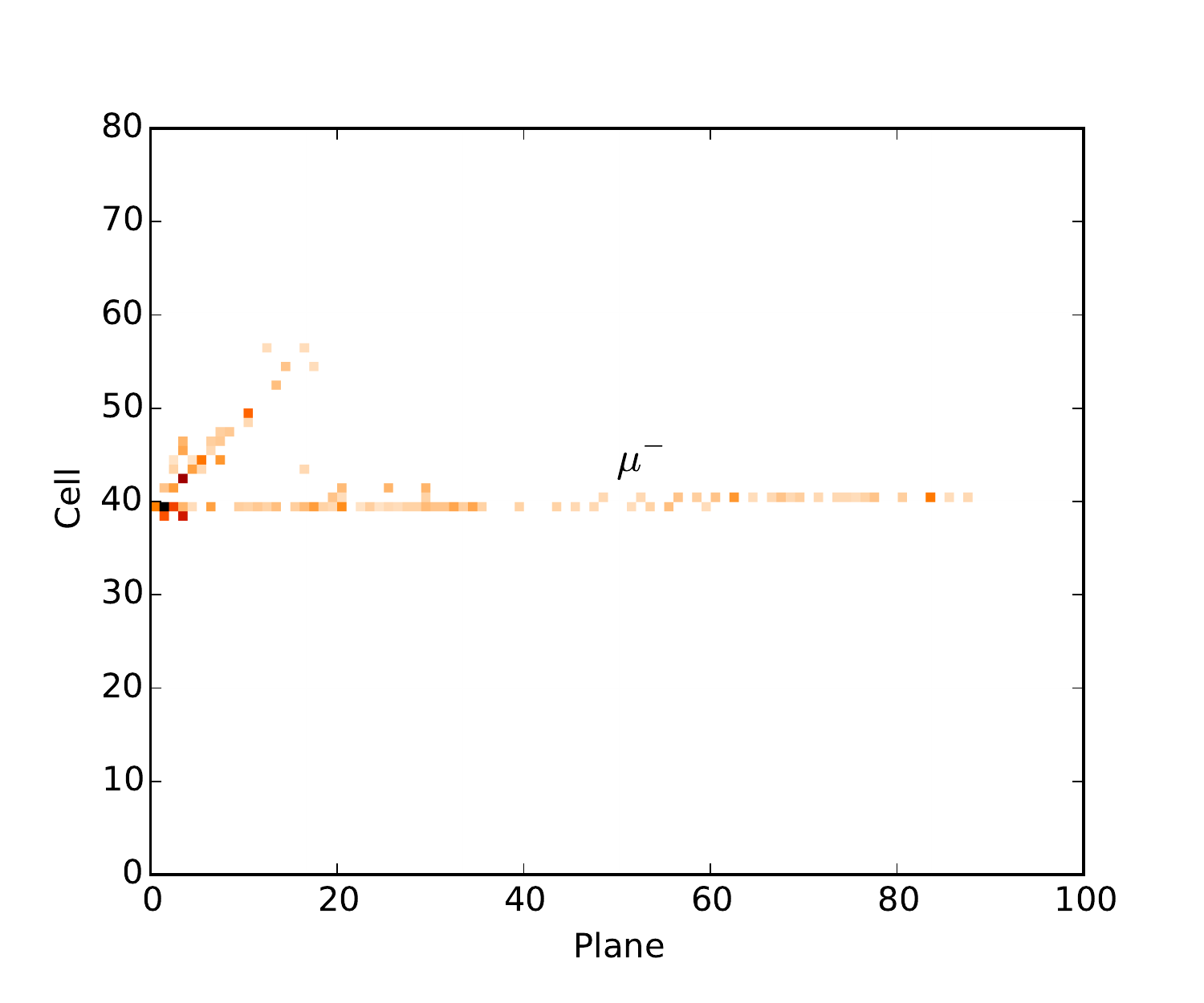}
          \vspace{-20pt}
          \caption*{\xview}
        \end{subfigure}
        \begin{subfigure}[c]{0.49\textwidth}
          \includegraphics[width=\textwidth]{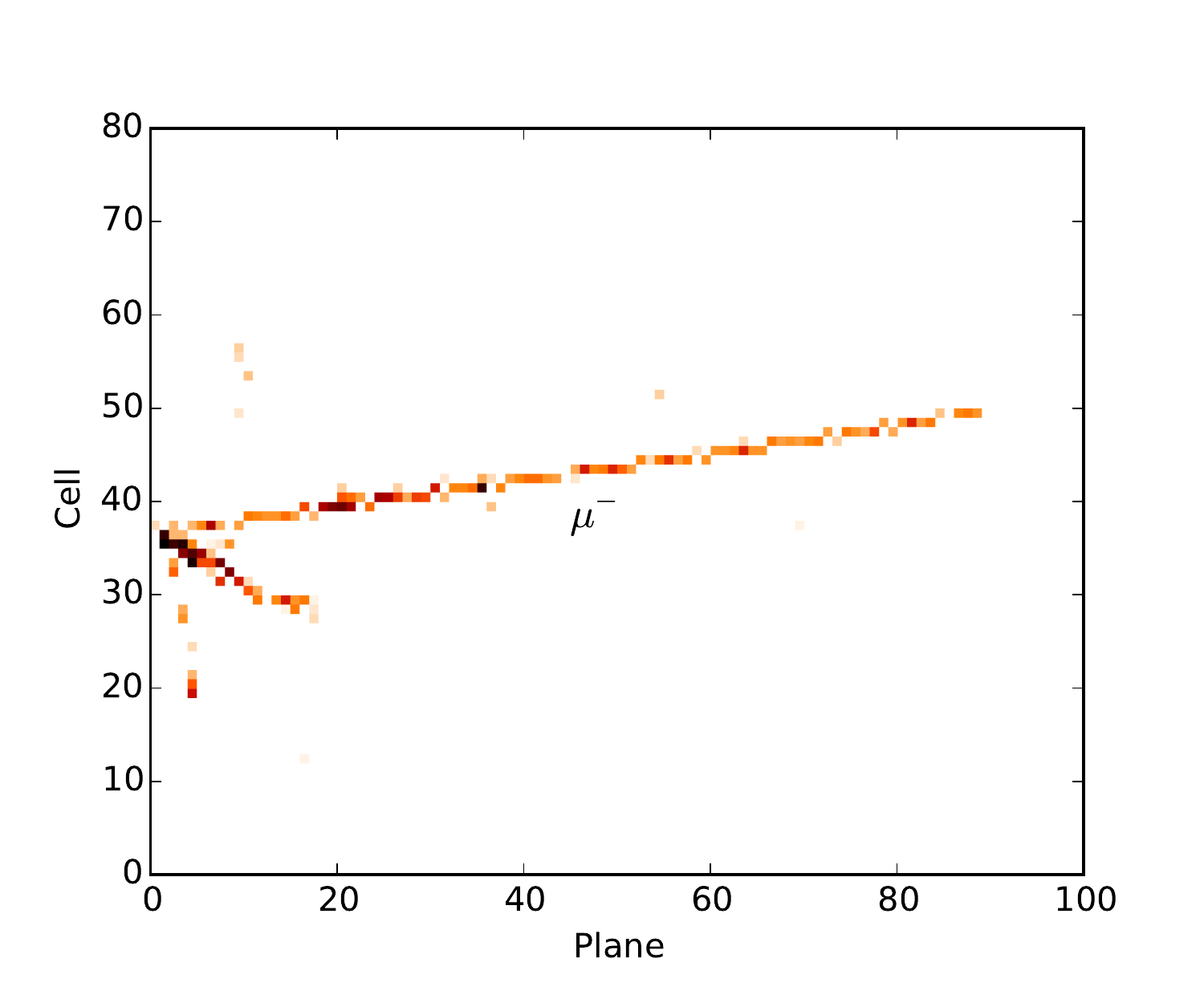}
          \vspace{-20pt}
          \caption*{\yview}
        \end{subfigure}
        \vspace{-10pt}
        \caption{\numu CC interaction.}
        \label{pixnumu}
      \end{center}
    \end{subfigure}
    \begin{subfigure}[c]{\textwidth}
      \begin{center}
        \begin{subfigure}[c]{0.49\textwidth}
          \includegraphics[width=\textwidth]{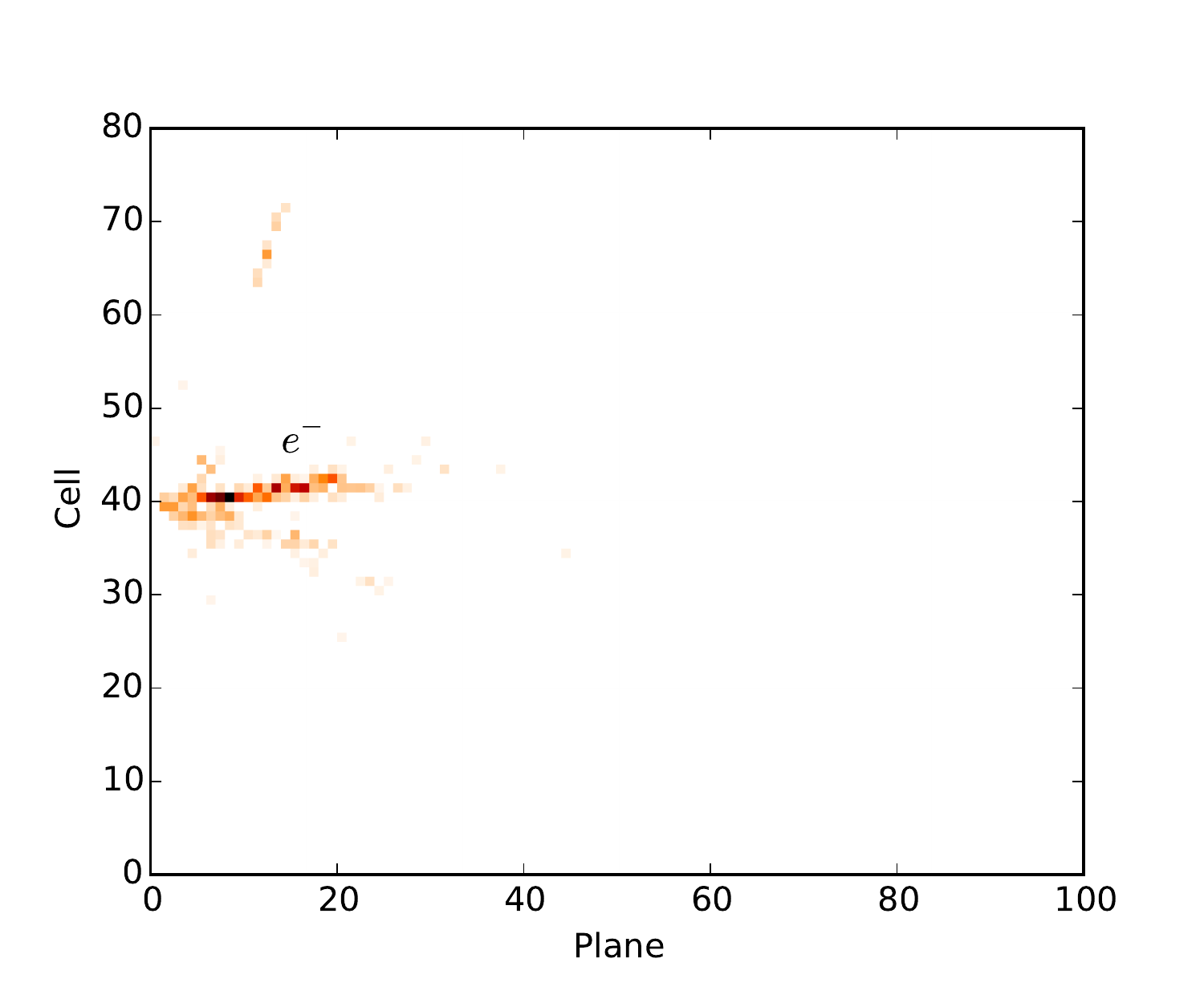}
          \vspace{-20pt}
          \caption*{\xview}
        \end{subfigure}
        \begin{subfigure}[c]{0.49\textwidth}
          \includegraphics[width=\textwidth]{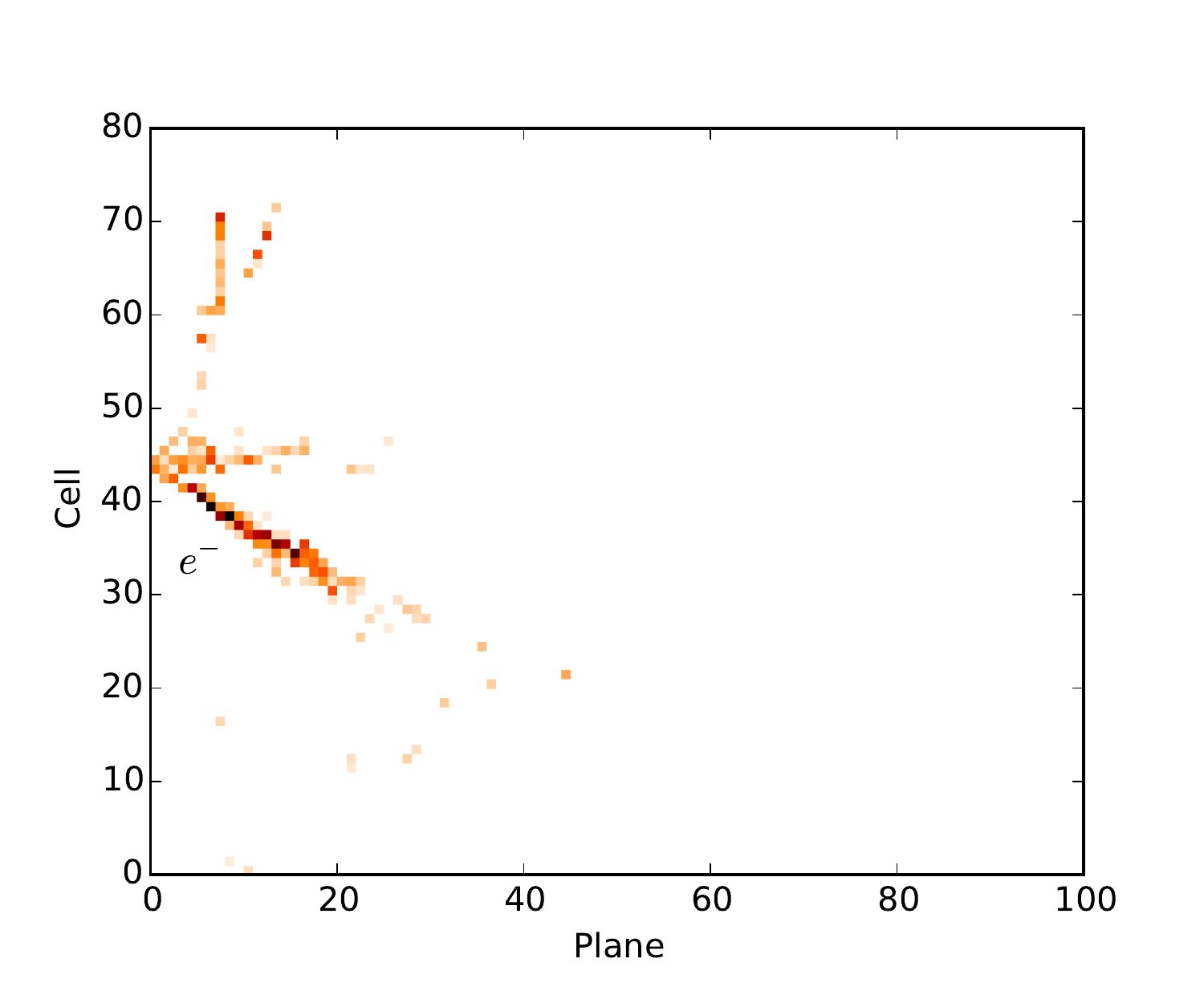}
          \vspace{-20pt}
          \caption*{\yview}
        \end{subfigure}
        \vspace{-10pt}
        \caption{\nue CC interaction.}
        \label{pixnue}
      \end{center}
    \end{subfigure}
    \begin{subfigure}[c]{\textwidth}
      \begin{center}
        \begin{subfigure}[c]{0.49\textwidth}
          \includegraphics[width=\textwidth]{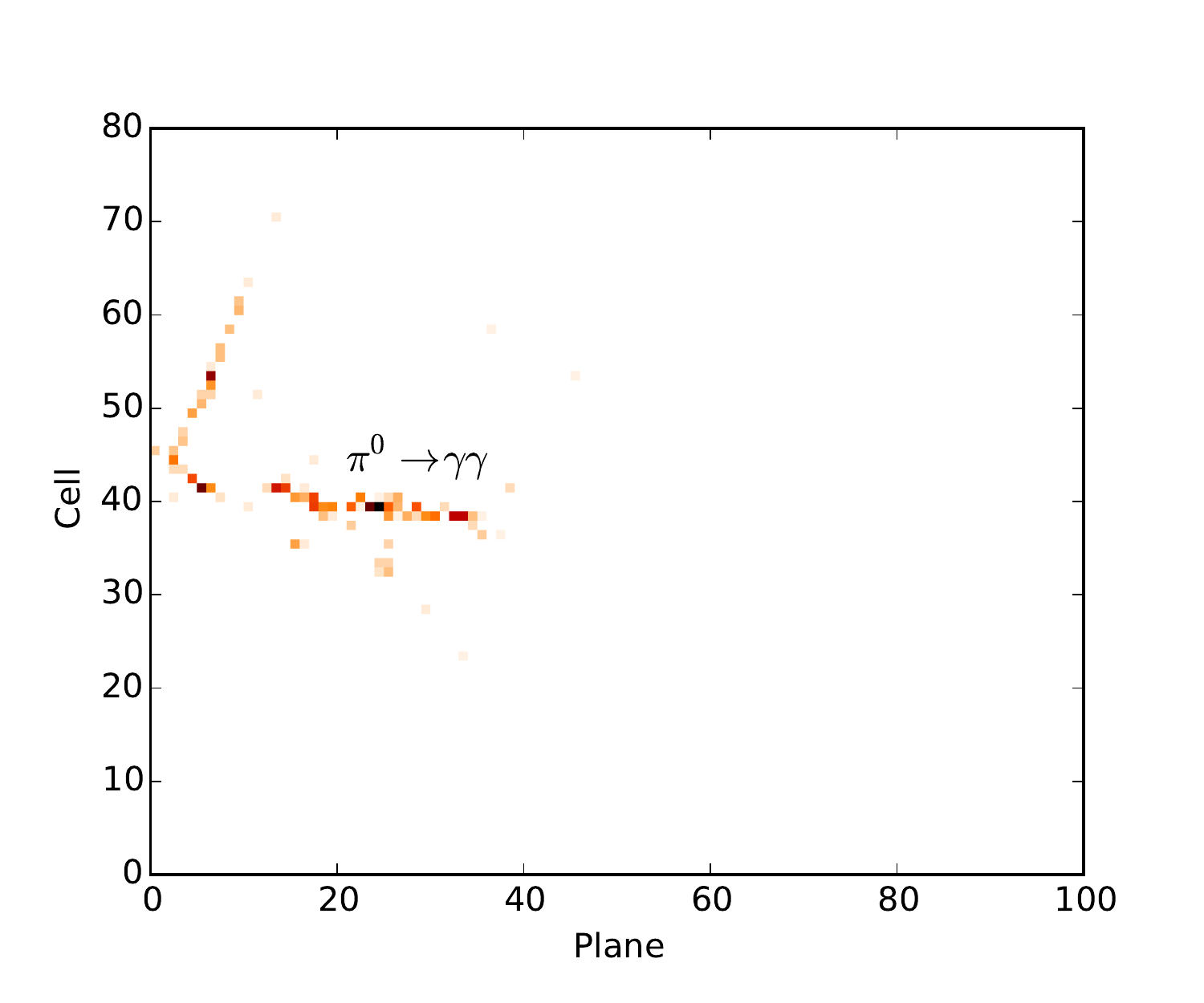}
          \vspace{-20pt}
          \caption*{\xview}
        \end{subfigure}
        \begin{subfigure}[c]{0.49\textwidth}
          \includegraphics[width=\textwidth]{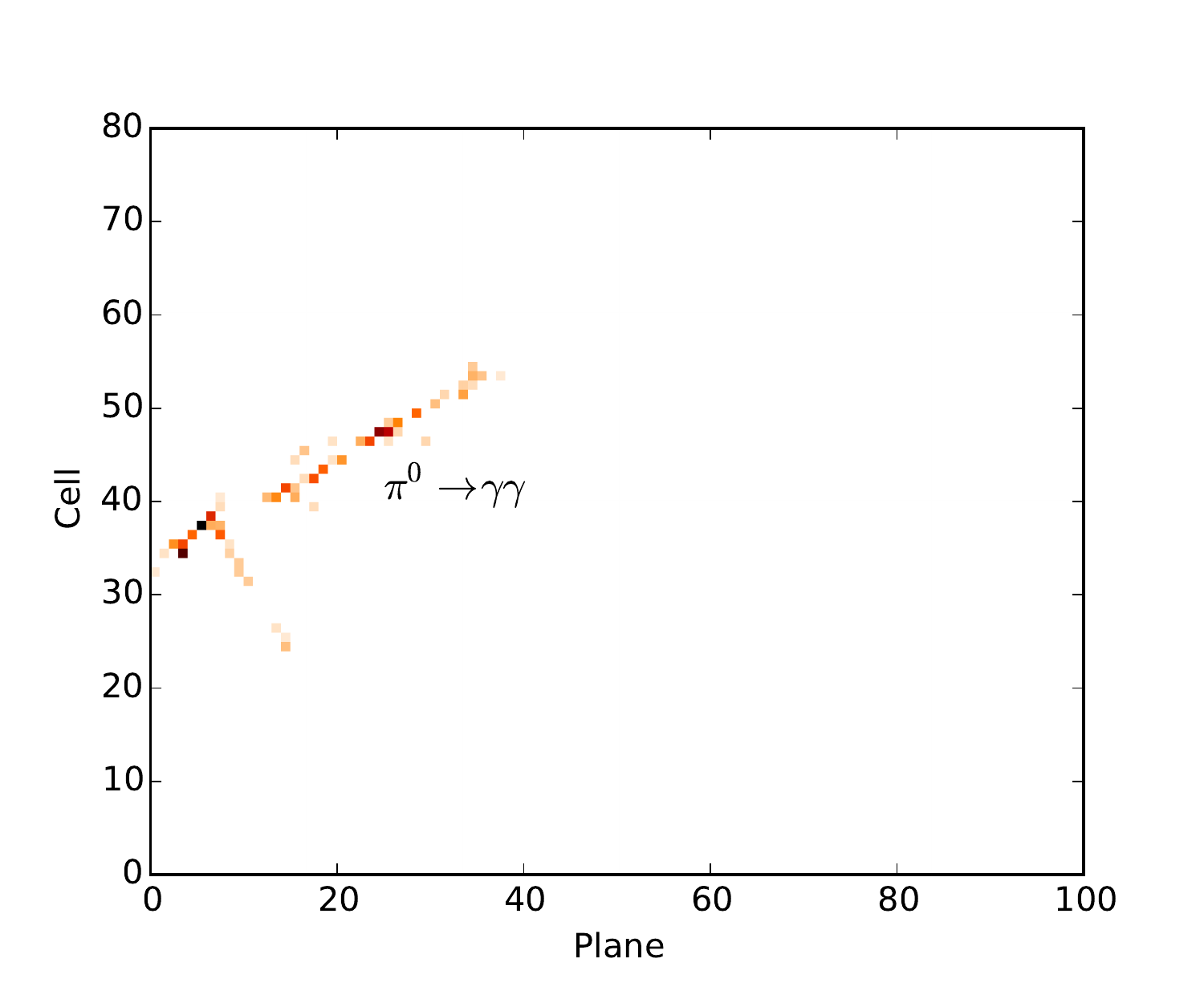}
          \vspace{-20pt}
          \caption*{\yview}
        \end{subfigure}
        \vspace{-10pt}
        \caption{NC interaction.}
        \label{pixnc}
      \end{center}
    \end{subfigure}
  \end{center}
  \caption{Example CNN image input}{
    Input given to the CNN for an
    example \numuCC interaction (top), \nueCC interaction
    (middle), and $\nu$ NC interaction (bottom). Hits in the X view of
    the NOvA detector are shown on the left, and hits in the Y view
    are shown on the right.}\label{pixelmap}
\end{figure*}

In order to optimize data storage and transfer in the training stage,
the pixel intensities were encoded using 8-bits which saturate for hits with energy above 278~MeV.  This conversion
offers a factor of eight savings over a floating point representation without significantly compromising the representational capacity, as shown in Figure~\ref{fig:CompressedEnergy}. These savings were especially important in reading data from disk into memory
during training. This map of 8-bit resolution hits, roughly analogous
to an image, was the input to our neural
network. Figure~\ref{pixelmap} shows examples of the input to the
neural network for three distinct neutrino interaction types.

\subsection*{CVN Architecture}

\begin{figure}[tbp]
\begin{center}
\includegraphics[height=0.8\textheight]{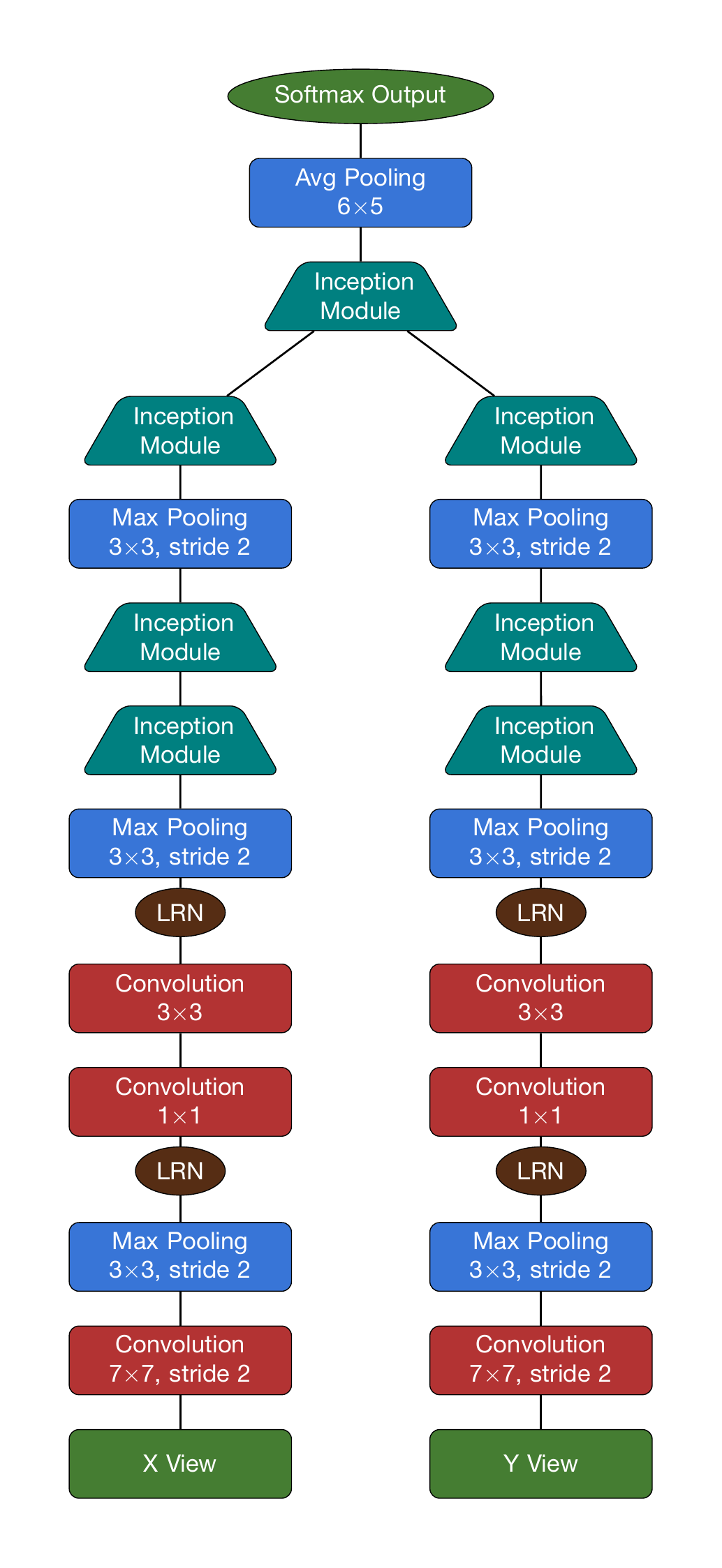}
\end{center}
\caption{Diagram of the CNN architecture used
  for event classification.}{Starting with the input at the bottom,
  the network has separate branches for the \xview and \yview.  Each
  branch undergoes successive convolution, pooling, and local response
  normalization (LRN).  Inception modules are used in downstream
  layers.  The two views are merged and passed through a final
  inception module, and pooled.  The output of the network comes from
  softmax units.  }
\label{arch}
\end{figure}

Our implementation of CVN was developed using the Caffe~\cite{caffe}
framework. Caffe is an open framework for deep learning applications
which is highly modular and makes accelerated training on graphics
processing units straightforward.  Common layer types are
pre-implemented in Caffe and can be arranged into new architectures by
specifying the desired layers and their connections in a configuration
file. Caffe is packaged with a configuration file implementing the
\googlenet architecture, and we used this as a starting point for
designing our own network which is shown in Figure~\ref{arch}.

The CVN architecture differs from the \googlenet architecture in a few
important ways. First, unlike most natural image problems, we have two
distinct views of the same image rather than a single image described
in multiple color channels.  In order to best extract the information
of each view, the channels corresponding to the X and Y-views were
split, and each resulting image was sent down a parallel architecture
based on \googlenet.  Since our images are simpler than natural
images, it was found that the nine inception modules used in the
\googlenet network did not improve our results; therefore, we
truncated the parallel \googlenet networks after three inception
modules. At that point, the features from each view were concatenated
together and passed through one final inception module to extract
combined features. The output of the final inception module was
down-sampled using an average pooling layer. During an evaluation
(forward) pass, classifier outputs were calculated using the
\textit{softmax} function or normalized exponential
function~\cite{Bishop:2006:PRM:1162264} such that the sum of the
classifier outputs was always equal to one.

\begin{figure*}
  \begin{center}
    \centering
    \begin{subfigure}[c]{0.5\textwidth}
      \includegraphics[width=\textwidth]{figures/view_truetype2_caltype2_event274_y.pdf}
    \end{subfigure}%
    ~
    \begin{subfigure}[c]{0.5\textwidth}
      \centering
      \includegraphics[width=\textwidth]{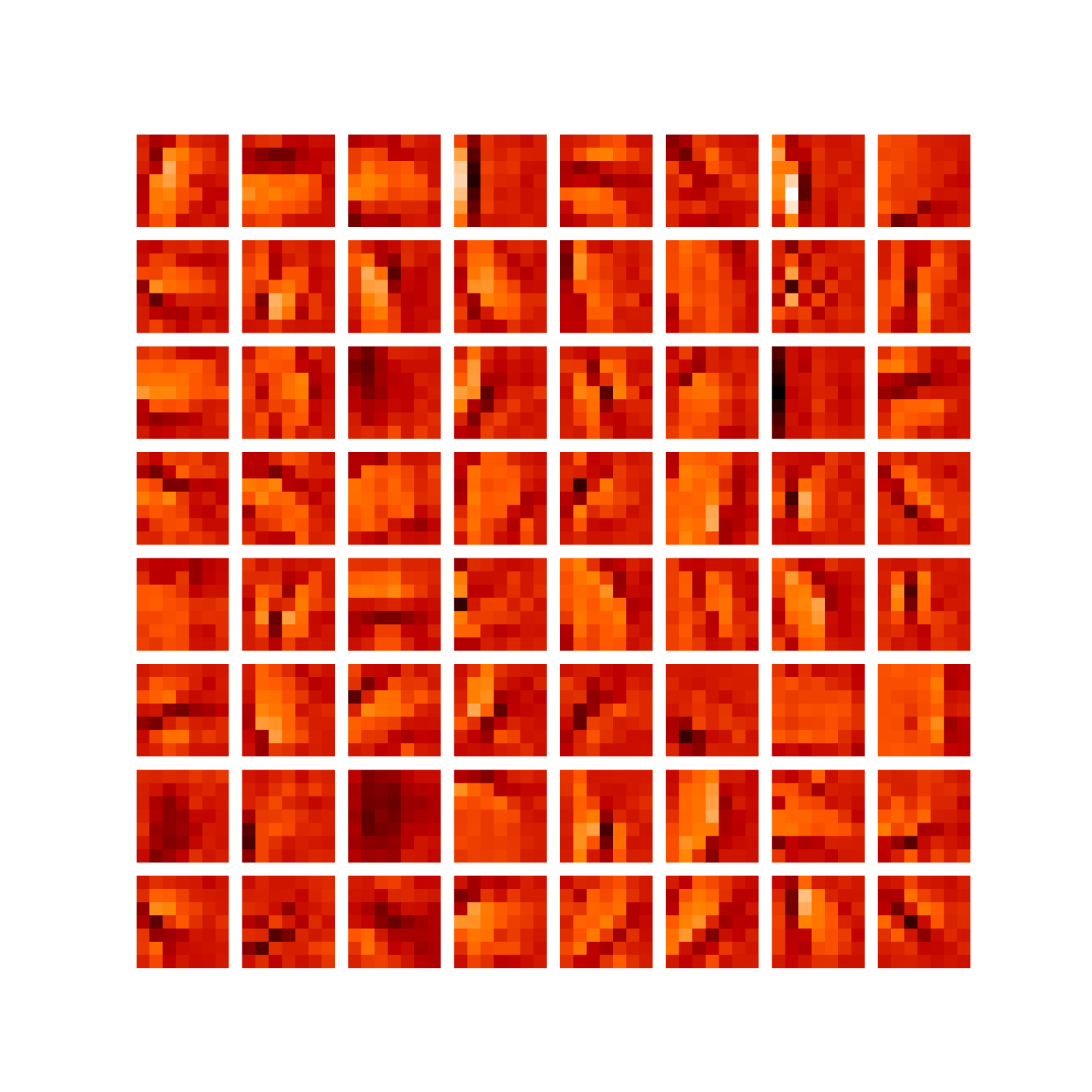}
    \end{subfigure}
    \begin{subfigure}[c]{.9\textwidth}
    \centering
    \vspace{30pt}
    \includegraphics[width=\textwidth]{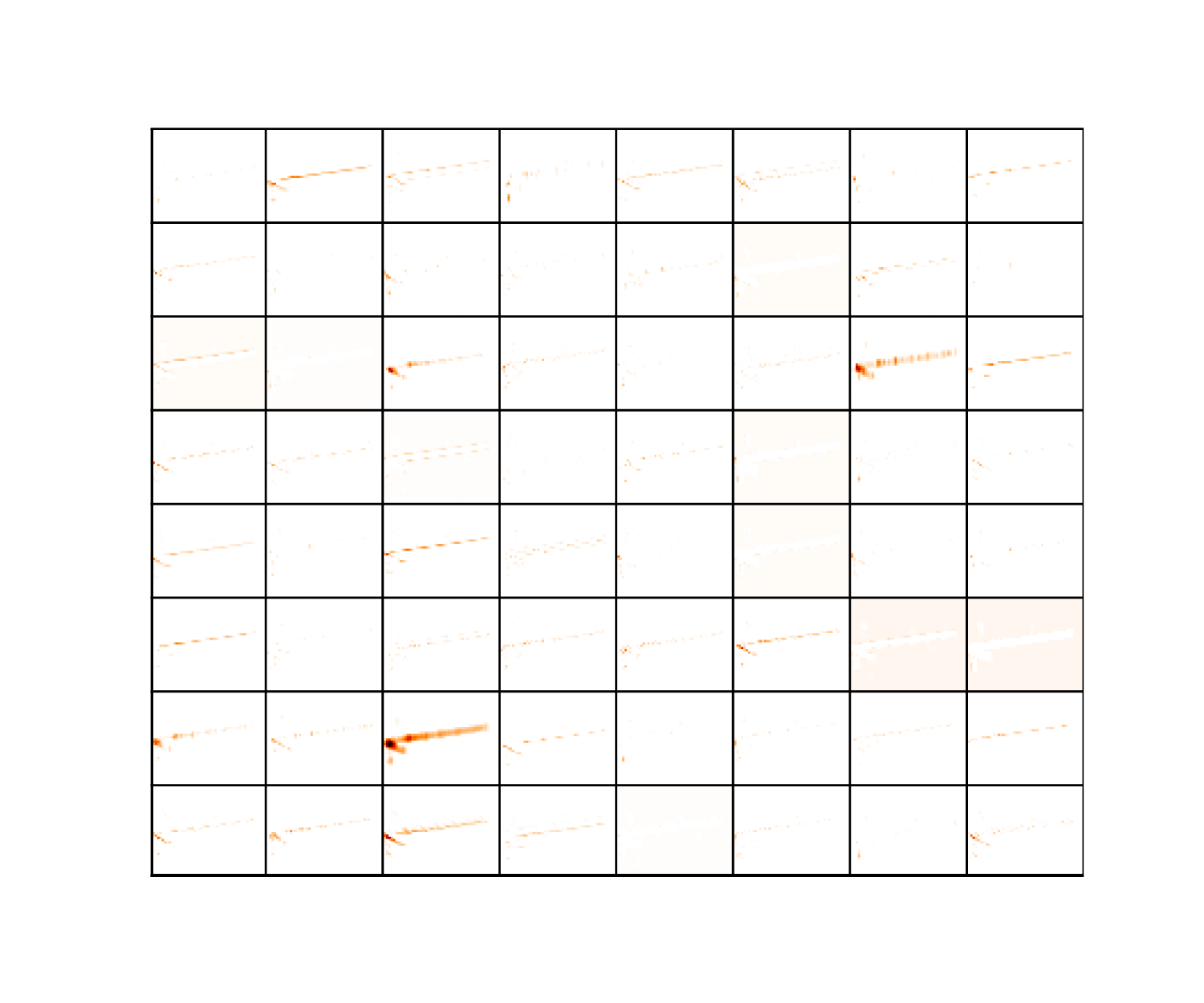}
  \end{subfigure}
\end{center}
\caption{$7\times7$ convolutional filter and example output}{
  This figure shows the Y view of an example \numu CC
  interaction (top left), all 64 convolutional filters from the
  early Y view $7\times7$ convolutional layer of our trained network (top
  right), and the output of applying those trained filters to the
  example event (bottom).}\label{featuremap}
\end{figure*}

\begin{figure*}
  \begin{center}
  \begin{subfigure}[b]{0.69\textwidth}
    \centering
    \includegraphics[width=\textwidth,viewport=10 13 170 115, clip=true]{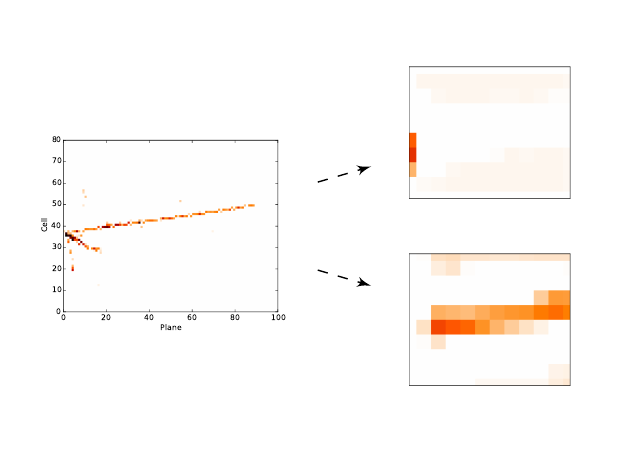}
  \end{subfigure}%
  \vspace{-10pt}

  \begin{subfigure}[b]{0.69\textwidth}
    \centering
    \includegraphics[width=\textwidth,viewport=10 13 170 115, clip=true]{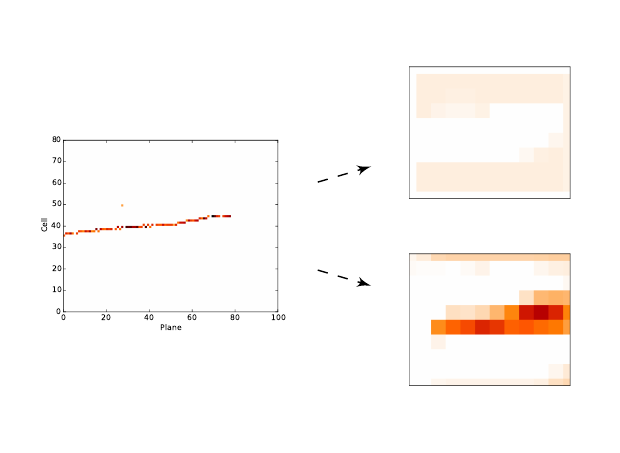}
  \end{subfigure}%
  \vspace{-10pt}

  \begin{subfigure}[b]{0.69\textwidth}
    \centering
    \includegraphics[width=\textwidth,viewport=10 13 170 115, clip=true]{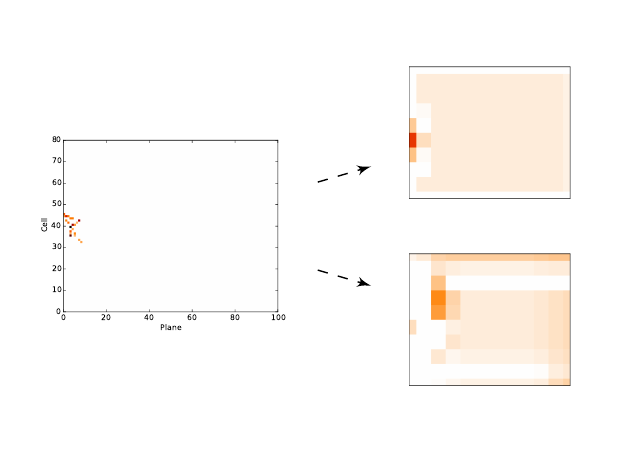}
  \end{subfigure}%
  \vspace{-15pt}
  \end{center}
  \caption{Output of the first inception module}
   {Shown above are three example input images
    and corresponding example human readable
    feature maps from the output of the first inception module in the
    Y view branch of our trained network. Darker regions indicate greater
    activation, and since this is the output from an early convolutional layer
    the regions correspond to the regions of the original image.
    The top-most feature map shows strong responses only in regions where hadronic
    activity can be found in the original image 
    and the bottom-most feature map shows strong activation only along the path of the muon track. 
    Shown are an example \numuCC DIS interaction (top),
    \numuCC QE interaction (middle), and $\nu$ NC interaction
    (bottom).}\label{featurexamples}
\end{figure*}

\subsection*{Training}

\begin{figure}[htb]
\begin{center}
\includegraphics[width=0.6\textwidth]{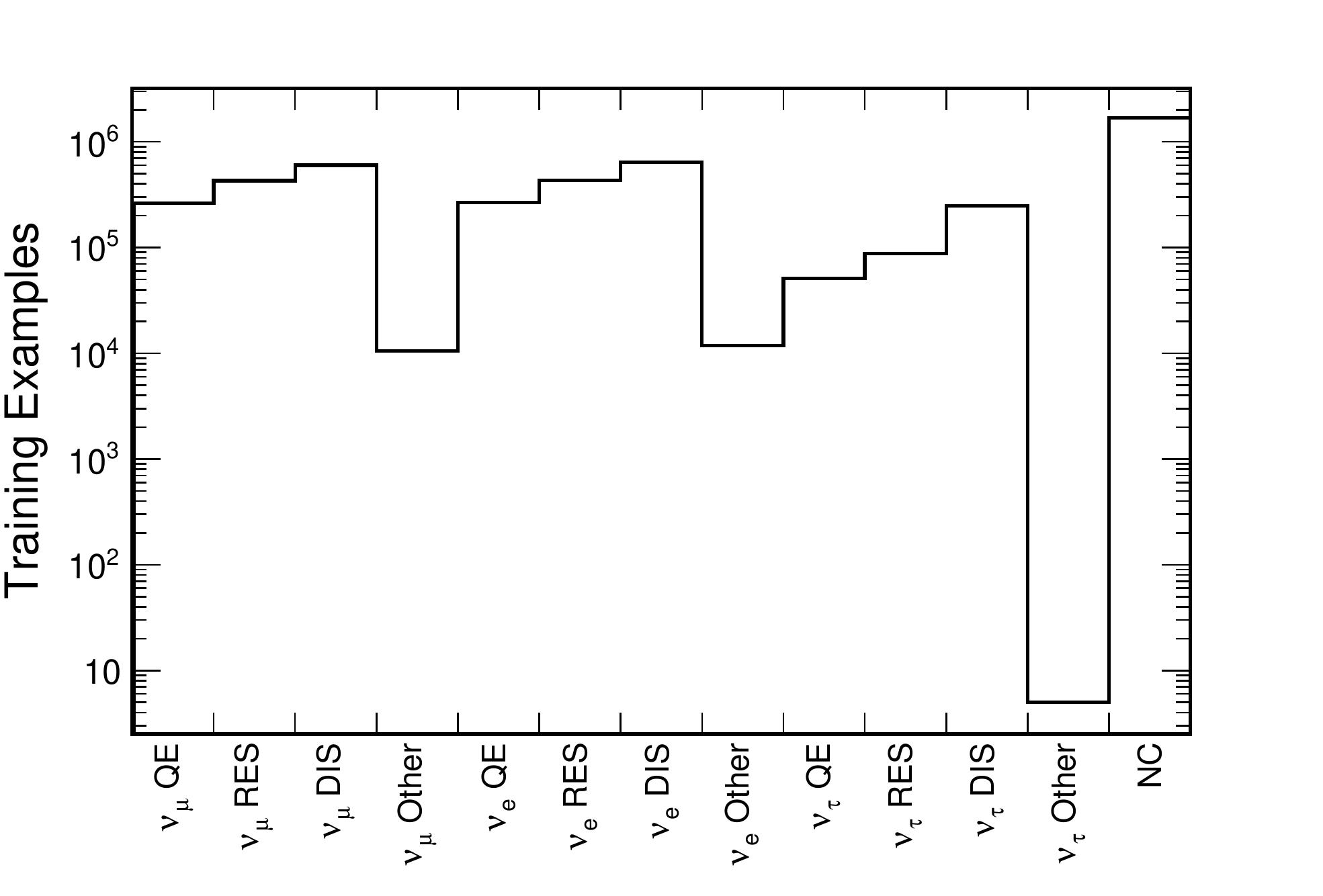}
\end{center}
\caption{\label{catfig} Number of events of each category in our training sample}
\end{figure}

A sample of 4.7 million simulated \nova Far Detector neutrino events
was used as the input to our training, with 80\% of the sample used for training and 20\% used for testing. FLUKA and
FLUGG~\cite{fluka1,fluka2} were used to simulate the flux in the NuMI
beamline~\cite{beamline}.  Before oscillations, the NuMI beam is composed mostly of \numu with 2.1\% intrinsic \nue contamination.  To model \nue and \nutau appearing through oscillation, we perform simulations where the \numu flux has been converted to another flavor. The training sample was composed of one third simulation with the expected, unoscillated flux, one third where the \numu flux was converted to \nue, and one third where the \numu flux was converted to \nutau. Neutrino-nucleus interactions were simulated using the
\genie~\cite{genie} package and GEANT4~\cite{geant1, geant2} was used
to propagate products of the neutrino interactions through a detailed
model of the \nova~detectors.  Custom NOvA simulation software
converted energy depositions into simulated electronic signals which
correspond to the detector output~\cite{novaSim}. The only
requirements placed on the training events were that they were
required to have 15 distinct hits.  Distinct categories were created
for each \genie CC interaction mode; $(\nu_\mu-{\rm CC},\nu_e-{\rm
  CC},\nu_\tau-{\rm CC}) \times (QE, RES, DIS, Other)$ and one
category was used for all \genie NC interactions to give the
distribution of events by label shown in Figure~\ref{catfig}.

We made use of a Mini-Batch~\cite{LeCun1998} training strategy which
simultaneously evaluated 32 training examples in each iteration.
During training the final softmax output was used to calculate a
multinomial logistic loss~\cite{GVK218647751} by comparing that output
to the true target value. We found that we converged to the most optimal versions of CVN by adopting a training strategy whereby the step
size of the stochastic gradient descent dropped at fixed intervals. These intervals were chosen to be at points where the network loss improvement rate had plateaued.
This strategy, in
combination with the vectorized implementation of mini-batch gradient
descent within the Caffe framework, allowed for rapid and accurate
optimization of our network.

To ensure that a trained model could reasonably generalize beyond the
training sample, we employed multiple regularization techniques. We
included penalty terms in the back-propagation calculation which are
proportional to the square of the weights, thus
constraining them to remain small during training.  Keeping the
weights small prevented the model from responding too strongly to any
one input feature. In addition, we applied the dropout technique with
$r = 0.4$ in the pooling step immediately before the final fully
connected layer.

Perhaps the most robust defense against over training is more training
data. We augmented our sample using two techniques to add variation to
the dataset.  First, pixel intensities were varied by adding Gaussian
noise with a standard deviation of 1\%. Adding noise had the benefit
of training the network to rely less heavily on the intensity in each
pixel and to adapt to the fluctuations encountered in real data.
Second, events were randomly selected to be reflected in the cell
dimension, which is roughly transverse to the beam direction. Symmetry
in the cell dimension is not perfect; the beam axis is directed
$3^{\circ}$ above detector horizon and attenuation in the optical
fiber causes thresholding to become more significant for hits further
from the readout electronics. However, these effects are small and
their presence in fact aids in enhancing variation of the training
sample and the robustness of the training against the precise details
of the simulation.

\begin{figure}[tbp]
\begin{center}
\includegraphics[width=0.8\textwidth]{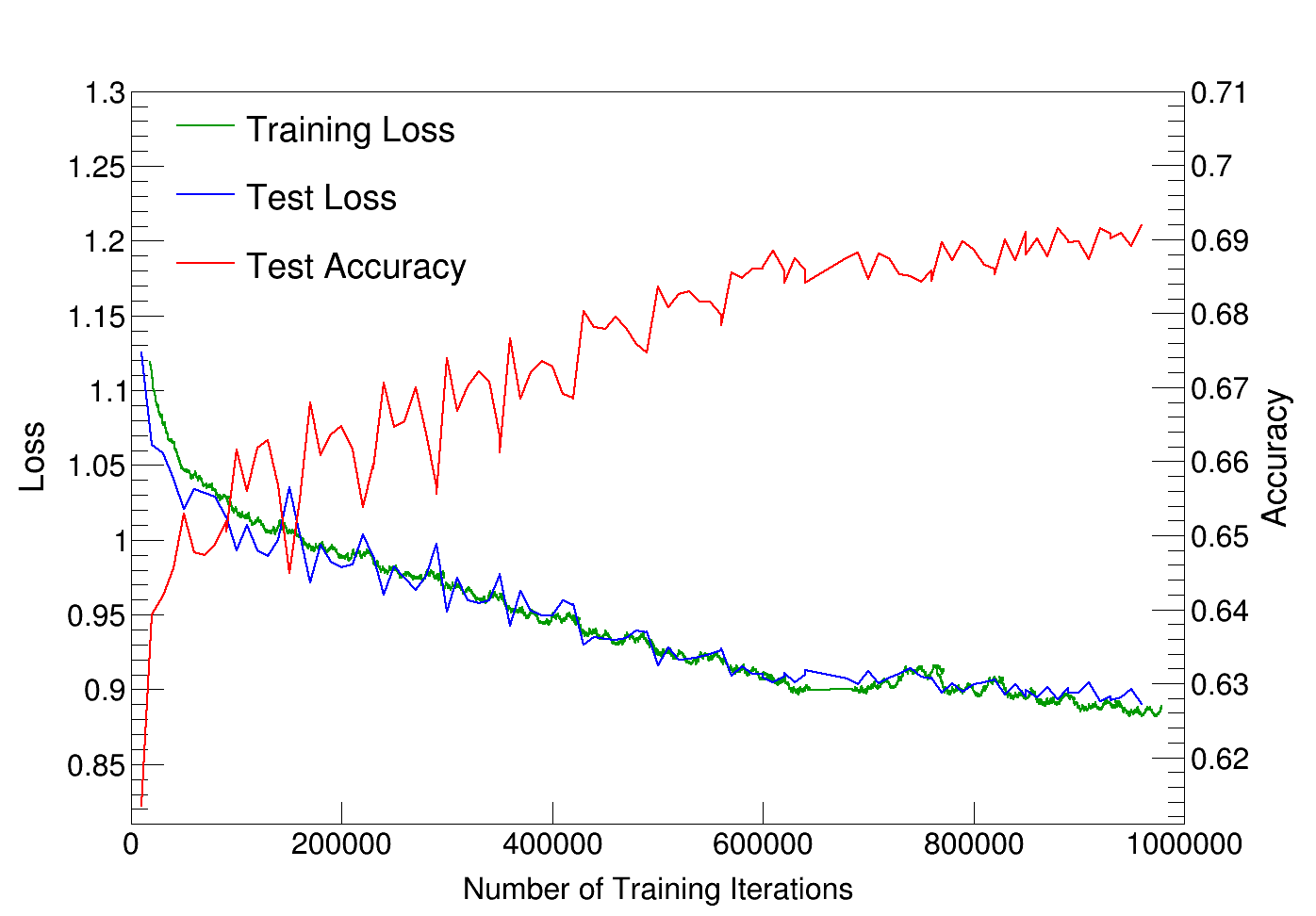}
\end{center}
\caption{Loss calculated on the training dataset (blue), loss calculated on the test dataset (green), and accuracy calculated on the test dataset (red) as a function of the number of training iterations the network has seen. The training loss shown here is a rolling window average of 40,000 iterations, such that each point represents the average over 128,000 training examples. Each test iteration is the average loss over 256,000 test examples.}
\label{lossovertime}
\end{figure}

The training of the network presented here was carried out on a pair of
NVIDIA Tesla K40s~\cite{cuda} over a full week of GPU hours. Figure \ref{lossovertime} shows loss
calculated for our training and test samples as a function of number of
training iterations; the extremely similar loss on both samples throughout 
training is a strong indication that the network has not overtrained.

\section{Results} \label{sec:results}

The same \nova simulation and software that allowed us to build our
training and test samples also allowed us to readily assess the
performance of our identification algorithm.  To measure performance
we used a statistically independent sample of neutrino interactions
from that used in the training and testing of CVN.  The sample was
weighted by the simulated \nova flux and by neutrino oscillation
probabilities from~\cite{pdg} using the \nova baseline~\cite{nova} and
matter density calculated from the CRUST 2.0 model of the Earth's
crust~\cite{crust} in order to create a representative mixture of
different beam events in the detector. The $\nu_{e}$ selection test
uses the preselection described in~\cite{Adamson:2016tbq}, and the
$\nu_{\mu}$ selection uses the preselection described
in~\cite{Adamson:2016xxw}. These preselections are designed to reject
cosmic backgrounds while retaining well-contained neutrino events
inside the signal energy window with high efficiency. We quote our
selection efficiencies relative to true contained signal, again
matching the approach described in ~\cite{Adamson:2016tbq} for
$\nu_{e}$ and ~\cite{Adamson:2016xxw} for $\nu_{\mu}$ tests
respectively.

\begin{figure*}
  \begin{center}
  \begin{subfigure}[b]{0.9\textwidth}
    \centering
    \includegraphics[width=\textwidth]{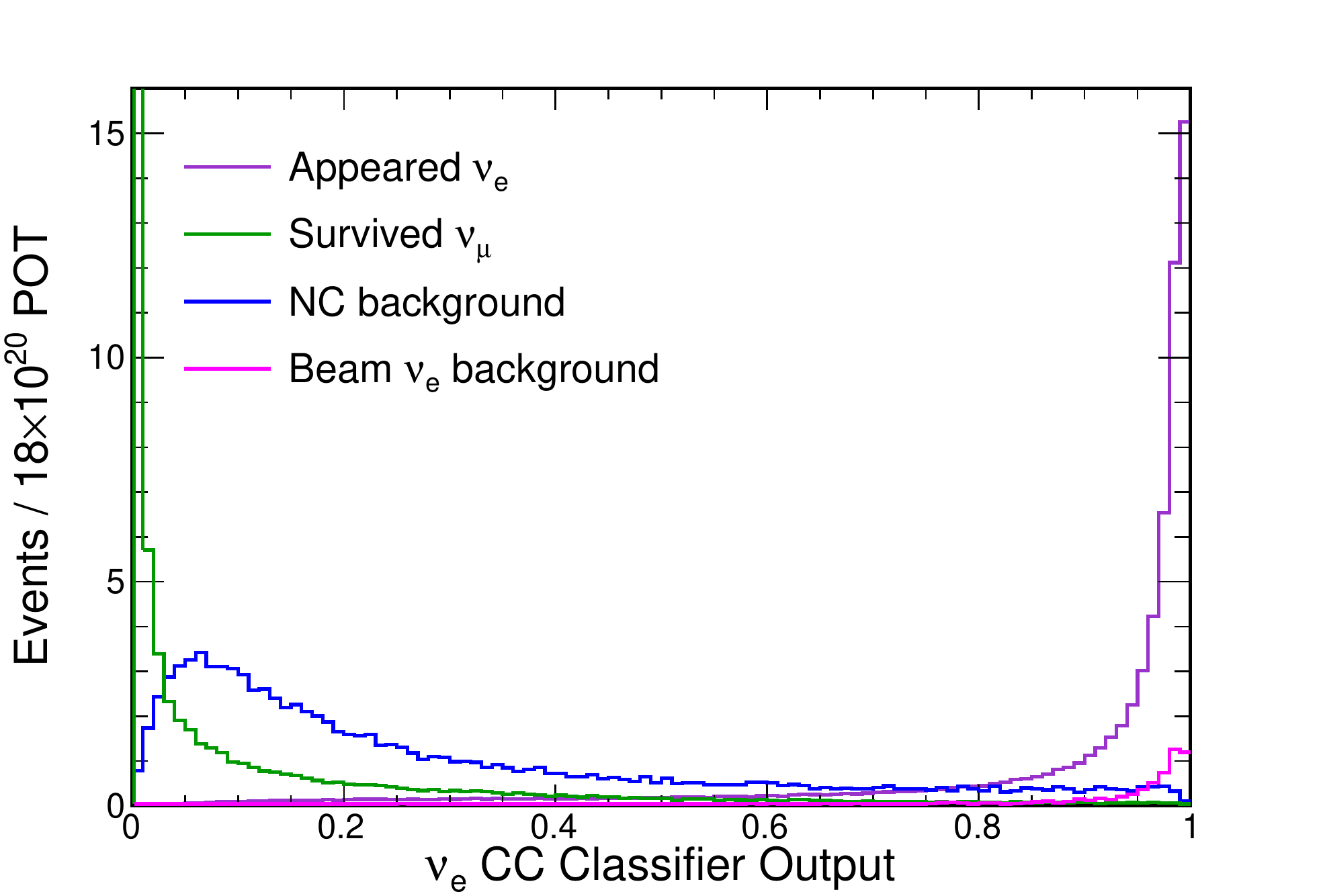}
    
  \end{subfigure}
  ~ 
  \begin{subfigure}[b]{0.9\textwidth}
    \centering
    \includegraphics[width=\textwidth]{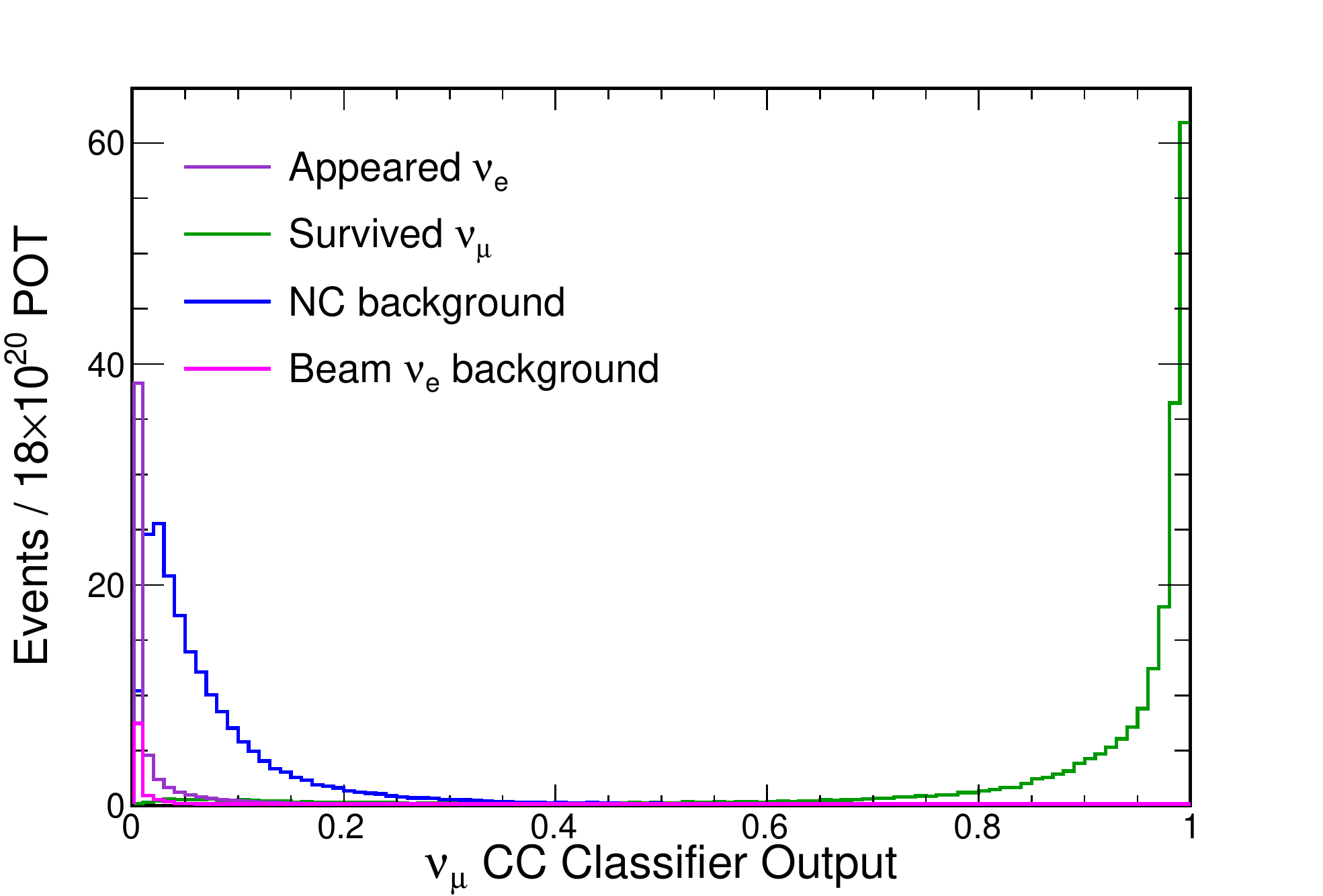}
    
  \end{subfigure}
  \end{center}

  \caption{\nueCC and \numuCC classifier output}
  {Distribution in $\nu_e$ CC Classifier output (top) and
    \numuCC Classifier output (bottom) for appearing electron
    neutrino CC interactions (violet), surviving CC
    muon neutrino (green), NC interactions (blue),
    and NuMI beam electron neutrino CC interactions
    (magenta).  The y-axis is truncated in the top figure such that the background is not
    fully visible in order to better show the signal distribution.
    Distributions are scaled to a NuMI exposure of $18\times 10^{20}$
    protons on target, full 14-kton Far Detector.}\label{pid}
\end{figure*}

\begin{figure*}
  \begin{center}
  \begin{subfigure}[b]{0.9\textwidth}
    \centering
    \includegraphics[width=\textwidth]{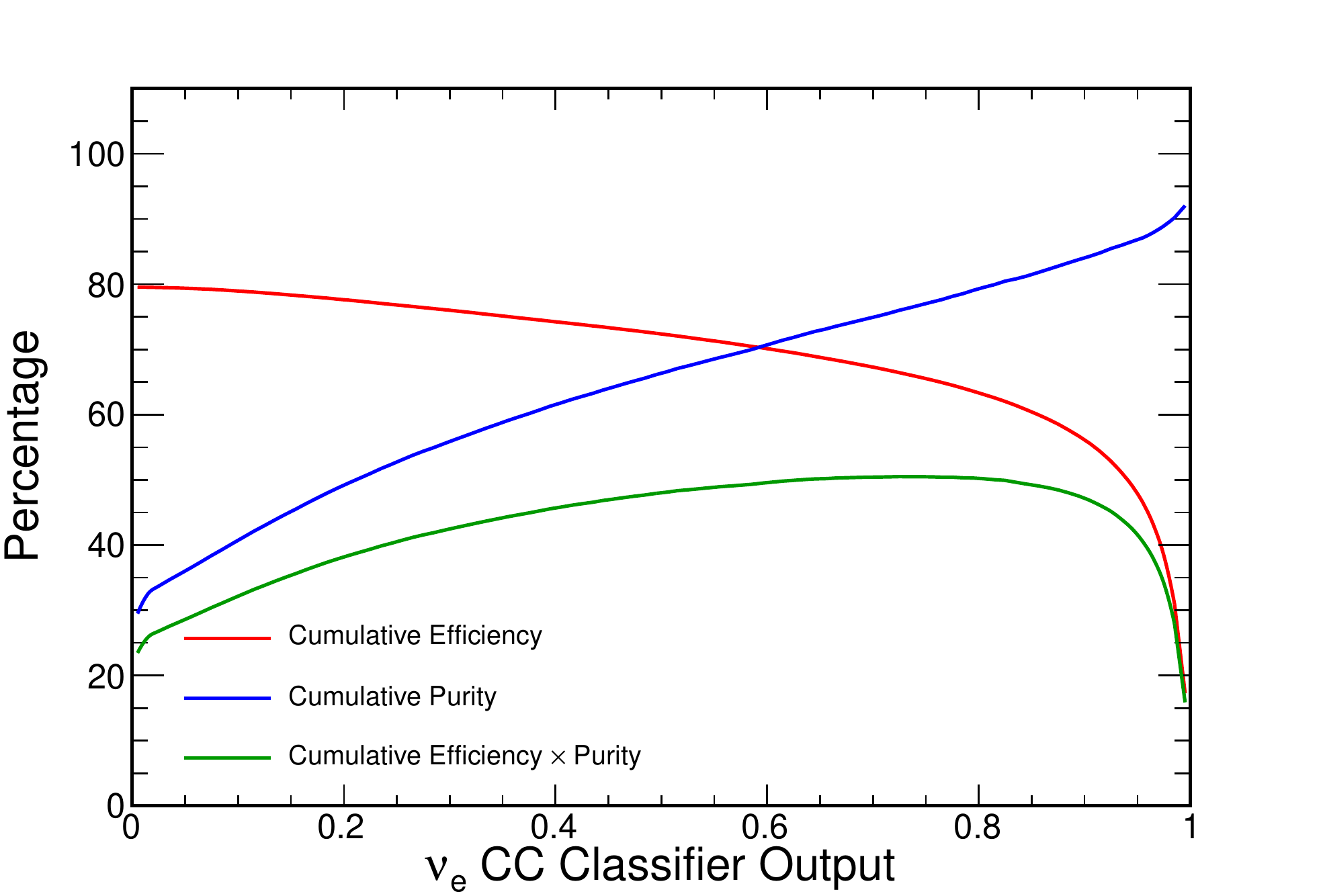}
    
  \end{subfigure}
  ~ 
  \begin{subfigure}[b]{0.9\textwidth}
    \centering
    \includegraphics[width=\textwidth]{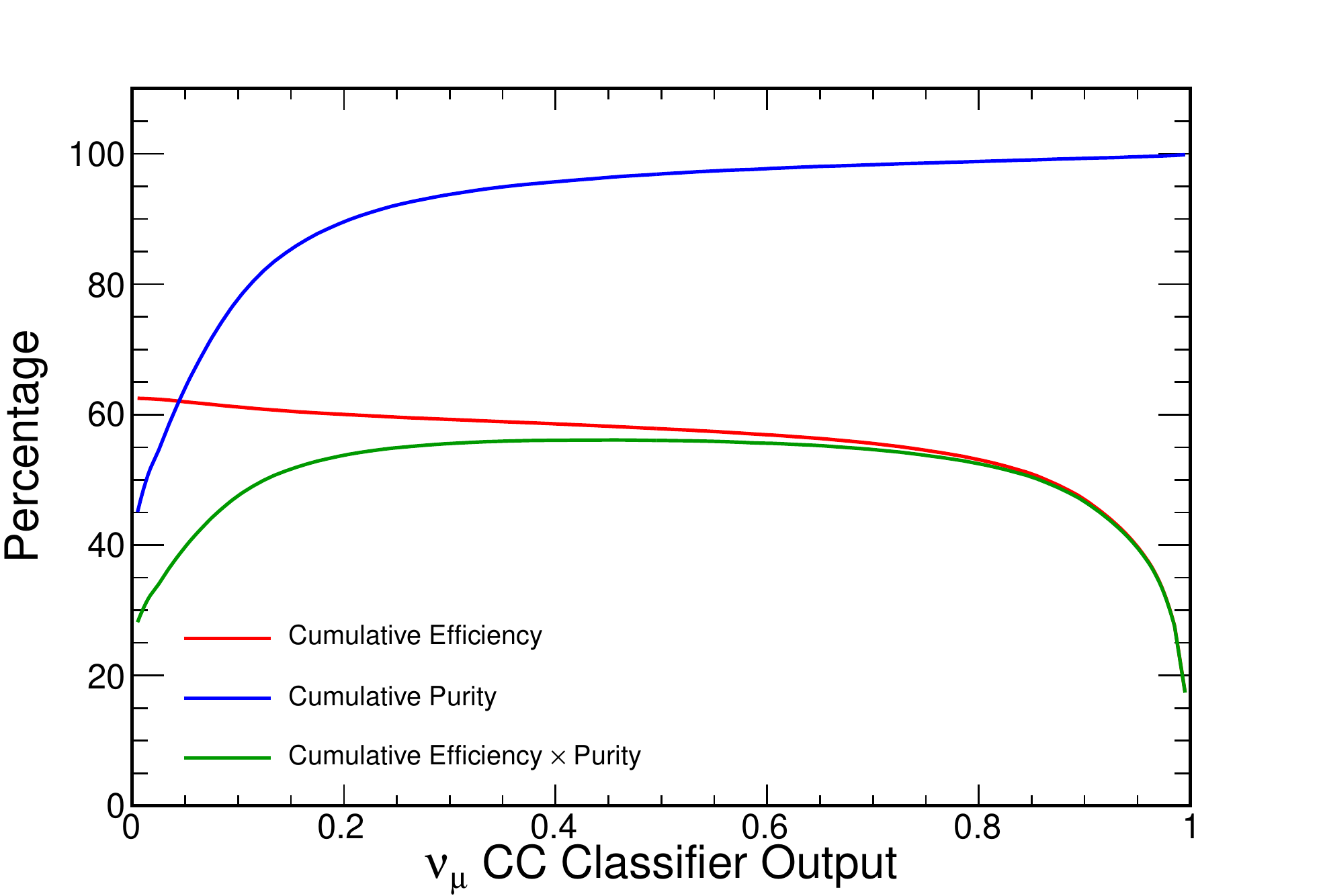}
    
  \end{subfigure}
  \end{center}

  \caption{\nueCC and \numuCC classifier output performance}
          {Cumulative efficiency (blue), purity (red), and their
            product (green) for \nueCC Classifier output (top) and
            \numuCC Classifier output (bottom). Efficiency is quoted
            relative to the true contained signal.}\label{effxpur}
\end{figure*}

Since the output of the final softmax layer in CVN is normalized to
one, it can be loosely interpreted as a probability of the input event
falling in each of the thirteen training categories.  For the results
presented in this paper a \nueCC classifier was derived from the sum
of the four \nue CC component probabilities.  Similarly, the four
\numuCC components were summed to yield a \numuCC classification.
Figure~\ref{pid} shows the distribution of the CVN \nueCC
classification parameter for true \nueCC events from $\nu_{\mu}
\rightarrow \nu_e$ oscillation and the various NuMI beam backgrounds
broken down by type.  Figure ~\ref{effxpur} shows the cumulative
efficiency, purity, and their product when selecting all events above
a particular CVN \nueCC classification parameter value.  Excellent
separation between signal and background is achieved such that the
only significant background remaining is that of electron neutrinos
present in the beam before oscillation; CVN does not attempt to
differentiate between \nueCC events from $\nu_{\mu} \rightarrow \nu_e$
oscillation and those from $\nu_e$ which are produced promptly in the
neutrino beam; these differ only in their energy
distributions. Figures ~\ref{pid} and ~\ref{effxpur} also show the
performance of the CVN \numuCC classification parameter. As with
$\nu_{e}$, excellent separation is achieved.

A common way to assess the performance of a signal selection is to
compute a Figure of Merit (FOM) given the number of selected signal
events $S$ and background events $B$. The FOM $=S/\sqrt{B}$ optimizes
for a pure sample useful for establishing the presence of the signal
$S$ in the presence of the background, while FOM $=S/\sqrt{S+B}$
optimizes for an efficient sample useful for making parameter
measurements with the signal $S$.  Table~\ref{fomTable} shows the
efficiency, purity, and event count at the maximal point for both
optimizations when using CVN to select \nueCC events, and for 
\numuCC events.  Using CVN we were able to set selection criteria well 
optimized for either FOM when searching for both surviving 
$\nu_{\mu}$ and appearing $\nu_{e}$ events.

\begin {table}[ht]
  \begin{subtable}[t]{1.\linewidth}
    {
      {
        \begin {center}
          \resizebox{\columnwidth}{!}{%
          \begin {tabular} {|c | c | c| c| c|c| c| c| c| c| c|}
            \hline
            & CVN Selection Value & $\nu_e$ sig & Tot bkg & NC & \numuCC & Beam $\nu_e$ & Signal Efficiency & Purity\\ \hline
            Contained Events & $-$ &  88.4 &         509.0 &        344.8 &        132.1 &        32.1 &         $-$ &      14.8\%\\
            $s/\sqrt{b}$ opt &       0.94 &  43.4 &         6.7 &  2.1 &  0.4 &  4.3 &  49.1\%&       86.6\%\\
            $s/\sqrt{s+b}$ opt &     0.72 &  58.8 &         18.6 &         10.3 &         2.1 &  6.1 &  66.4\%&       76.0\%\\ \hline            
          \end {tabular}
          }
        \end {center}
      }
    }
    \vskip-0.2in
    \caption{}

    {
      {
        \begin {center}
          \resizebox{\columnwidth}{!}{%
            \begin {tabular} {|c | c | c| c| c|c| c| c| c| c| c|}
              \hline
            & CVN Selection Value & $\nu_{\mu}$ sig & Tot bkg & NC & Appeared $\nu_e$ & Beam $\nu_e$ & Signal Efficiency & Purity\\ \hline
              Contained Events  & $-$ & 355.5 & 1269.8 & 1099.7 & 135.7 & 34.4 & $-$ & 21.9\%\\
              $s/\sqrt{b}$ opt & 0.99 & 61.8 & 0.1 & 0.1 & 0.0 & 0.0 & 17.4\%& 99.9\%\\
              $s/\sqrt{s+b}$ opt & 0.45 & 206.8 & 7.6 & 6.8 & 0.7 & 0.1 & 58.2\%& 96.4\%\\ \hline
            \end {tabular}
          }
        \end {center}
      }
    }
    \vskip-0.2in
    \caption {}

    \end{subtable}
    \caption {Tables showing relative selected event numbers for the
      various components of the NuMI beam, efficiency, and purity for two different optimizations for the
      selection of appearing electron neutrino CC
      interactions (a) and surviving muon neutrino CC interactions (b). Efficiency is shown here relative to the
      true contained signal. The numbers are scaled to an exposure of
      $18\times 10^{20}$ protons on target, full 14-kton Far Detector.}\label{fomTable}
  \end {table}

Perhaps the most important way to assess the performance of the CVN
classification parameters is to compare their performance to the
sophisticated identification algorithms already used in recent \nova
publications.  For \numuCC interactions the CVN measurement-optimized
efficiency of 58\% is comparable to the efficiency of 57\% quoted in
Ref.\cite{Adamson:2016xxw}.  This is a modest improvement but shows
that CVN algorithm does not underperform when used to identify a class
of events we expect to be particularly clear. For \nueCC interactions
the CVN signal-detection-optimized efficiency of 49\% is a significant
gain over the efficiency of 35\% quoted in in
Ref.\cite{Adamson:2016tbq}.  In both the \numuCC and \nueCC cases the
CVN purity very closely matches the purity of the samples reported
in~Refs.\cite{Adamson:2016xxw,Adamson:2016tbq}.  The \nueCC efficiency
improvement is significant not just because the \nueCC signal is
particularly hard to separate from its backgrounds, but additionally
because the measurements of \nue-appearance and the associated
parameters ($\theta_{13}$, neutrino mass hierarchy, etc.) are
statistics limited and hence significantly improved by increasing the
signal efficiency.

Tests of the CVN selectors response to
systematic uncertainties suggest no increased sensitivity to them compared to the 
the selectors used in~Refs.\cite{Adamson:2016xxw,Adamson:2016tbq}.
In particular we studied the variation in signal
and background selection for our \numuCC and \nueCC optimized cuts
for a subset of the dominant uncertainties in ~Refs.\cite{Adamson:2016xxw,Adamson:2016tbq},
specifically calibration and scintillator saturation uncertainties.
For \numuCC interactions the CVN measurement-optimized selection
signal sample was sensitive at the $1.9\%$ level, and the
extremely small background was sensitive at the $20\%$ level.
For \nueCC interactions the CVN signal-detection-optimized selection
signal sample was sensitive at the $1.6\%$ level, and the
extremely small background was sensitive at the $1.0\%$ level.
In both the \numuCC and \nueCC cases these shifts are small
and comparable to those seen in the selectors used in ~Refs.\cite{Adamson:2016xxw,Adamson:2016tbq}.

\section{Conclusion and future work} \label{sec:conclusions}

With minimal event reconstruction, we were able to build and train a
single algorithm which achieved excellent separation of signal and
background for both of the \nova electron-neutrino appearance and
muon-neutrino disappearance oscillation channels. This algorithm, CVN,
is a powerful approach to the problem of event classification and
represents a novel proof of concept -- that CNNs can work extremely
well with non-natural images like the readout of a sampling
calorimeter. We expect this approach to be transferable to a wide
range of detector technologies and analyses.

CVN also opens up other possibilities for event classification and
reconstruction which we have just started to explore. For example, the
same training and architecture reported here can be expanded to
include identification of $\nu$ NC and \nutauCC interactions which are
used by \nova in sterile neutrino and exotic physics searches. We also
tested the application of CVN to the problem of identification of
particles within neutrino events by presenting the algorithm with
images of individual, isolated particle clusters~\cite{Baird:2015pgm}
along with the images of the overall neutrino event. Initial results
have already shown improvements in the efficiency and purity in
particle identification over other likelihood methods employed by
\nova and may aid in future cross-section analyses. We are also
exploring improvements to cluster-finding and particle identification
using semantic segmentation~\cite{DBLP:journals/corr/LongSD14} which,
starting from the same minimal reconstruction used by CVN, may make it
possible to identify the particles which contributed to each ``pixel''
in the detector readout.

\acknowledgments

The authors thank the \nova collaboration for use of its Monte Carlo
simulation software and related tools.  We thank Gustav Larsson for
useful conversations, and we are grateful for Amitoj Singh and the
Fermilab Scientific Computing Division's efforts installing Caffe and
maintaining the GPU cluster used for training.

This work was supported by the US Department of Energy and the US
National Science Foundation. NOvA receives additional support from the Department of Science and Technology,
India; the European Research Council; the MSMT CR, Czech Republic; the
RAS, RMES, and RFBR, Russia; CNPq and FAPEG, Brazil; and the State and
University of Minnesota. We are grateful for the contributions of the
staff at the Ash River Laboratory, Argonne National Laboratory, and
Fermilab. Fermilab is operated by Fermi Research Alliance, LLC under
Contract No.~De-AC02-07CH11359 with the US DOE.

\bibliographystyle{JHEP}
\bibliography{cvn_paper}

\providecommand{\href}[2]{#2}\begingroup\raggedright\begin{thebibliography}{10}

\bibitem{knn}
N.~S. Altman, \emph{An introduction to kernel and nearest-neighbor
  nonparametric regression},
  \href{http://dx.doi.org/10.1080/00031305.1992.10475879}{\emph{The American
  Statistician} {\bf 46} (1992) 175--185},
  [\href{http://arxiv.org/abs/http://www.tandfonline.com/doi/pdf/10.1080/00031305.1992.10475879}{{[\tt
  http://www.tandfonline.com/doi/pdf/10.1080/00031305.1992.10475879]}}].

\bibitem{friedman2002stochastic}
J.~H. Friedman, \emph{Stochastic gradient boosting}, {\emph{Computational
  Statistics \& Data Analysis} {\bf 38} (2002) 367--378}.

\bibitem{Rosenblatt1961}
F.~Rosenblatt, \emph{Principles of Neurodynamics: Perceptrons and the Theory of
  Brain Mechanisms}.
\newblock Spartan Books, 1961.

\bibitem{reed1999neural}
R.~Reed and R.~Marks, \emph{Neural Smithing: Supervised Learning in Feedforward
  Artificial Neural Networks}.
\newblock A Bradford book. MIT Press, 1999.

\bibitem{LeCun1989}
Y.~LeCun, B.~Boser, J.~S. Denker, D.~Henderson, R.~E. Howard, W.~Hubbard
  et~al., \emph{Backpropagation applied to handwritten zip code recognition},
  \href{http://dx.doi.org/10.1162/neco.1989.1.4.541}{\emph{Neural Comput.} {\bf
  1} (Dec., 1989) 541--551}.

\bibitem{nova}
{\scshape NOvA} collaboration, D.~S. Ayres et~al., \emph{The {NOvA} technical
  design report},  FERMILAB-DESIGN-2007-01.

\bibitem{minerva}
{\scshape MINERvA} collaboration, L.~Aliaga et~al., \emph{Design, calibration,
  and performance of the {MINERvA} detector},
  \href{http://dx.doi.org/http://dx.doi.org/10.1016/j.nima.2013.12.053}{\emph{Nucl.
  Instrum. Meth. A} {\bf 743} (2014) 130 -- 159}.

\bibitem{minos}
{\scshape MINOS} collaboration, D.~G. Michael et~al., \emph{The magnetized
  steel and scintillator calorimeters of the {MINOS} experiment}, {\emph{Nucl.
  Instrum. Meth. A} {\bf 596} (2008) 190--228} FERMILAB-PUB-08-126,
  [\href{http://arxiv.org/abs/arXiv:0805.3170}{{[\tt arXiv:0805.3170]}}].

\bibitem{Bettini:1991fh}
A.~Bettini et~al., \emph{{The ICARUS liquid argon TPC: A Complete imaging
  device for particle physics}},
  \href{http://dx.doi.org/10.1016/0168-9002(92)90707-B}{\emph{Nucl. Instrum.
  Meth.} {\bf A315} (1992) 223--228}.

\bibitem{microboone}
H.~Chen et~al., \emph{Proposal for a new experiment using the booster and
  {NuMI} neutrino beamlines: {MicroBooNE}},  FERMILAB-PROPOSAL-0974.

\bibitem{dune1}
{\scshape DUNE} collaboration, R.~Acciarri et~al., \emph{{Long-Baseline
  Neutrino Facility (LBNF) and Deep Underground Neutrino Experiment (DUNE)
  Conceptual Design Report Volume 1: The LBNF and DUNE Projects}},
  \href{http://arxiv.org/abs/arXiv:1601.05471}{{[\tt arXiv:1601.05471]}}
  FERMILAB-DESIGN-2016-01, [\href{http://arxiv.org/abs/arXiv:1601.05471}{{[\tt
  arXiv:1601.05471]}}].

\bibitem{dune2}
{\scshape DUNE} collaboration, R.~Acciarri et~al., \emph{Long-baseline neutrino
  facility ({LBNF}) and deep underground neutrino experiment ({DUNE})
  conceptual design report volume 2: The physics program for {DUNE} at {LBNF}},
   \href{http://arxiv.org/abs/arXiv:1512.06148}{{[\tt arXiv:1512.06148]}}
  FERMILAB-DESIGN-2016-02, [\href{http://arxiv.org/abs/arXiv:1512.06148}{{[\tt
  arXiv:1512.06148]}}].

\bibitem{dune3}
{\scshape DUNE} collaboration, J.~Strait et~al., \emph{Long-baseline neutrino
  facility ({LBNF}) and deep underground neutrino experiment ({DUNE})
  conceptual design report volume 3: Long-baseline neutrino facility for
  {DUNE}},  \href{http://arxiv.org/abs/arXiv:1601.05823}{{[\tt
  arXiv:1601.05823]}} FERMILAB-DESIGN-2016-03,
  [\href{http://arxiv.org/abs/arXiv:1601.05823}{{[\tt arXiv:1601.05823]}}].

\bibitem{dune4}
{\scshape DUNE} collaboration, R.~Acciarri et~al., \emph{Long-baseline neutrino
  facility ({LBNF}) and deep underground neutrino experiment ({DUNE})
  conceptual design report, volume 4 the {DUNE} detectors at {LBNF}},
  \href{http://arxiv.org/abs/arXiv:1601.02984}{{[\tt arXiv:1601.02984]}}
  FERMILAB-DESIGN-2016-04, [\href{http://arxiv.org/abs/arXiv:1601.02984}{{[\tt
  arXiv:1601.02984]}}].

\bibitem{icecube1}
{\scshape IceCube} collaboration, J.~Ahrens et~al., \emph{Sensitivity of the
  {IceCube} detector to astrophysical sources of high energy muon neutrinos},
  \href{http://dx.doi.org/10.1016/j.astropartphys.2003.09.003}{\emph{Astropart.
  Phys.} {\bf 20} (2004) 507--532},
  [\href{http://arxiv.org/abs/astro-ph/0305196}{{[\tt astro-ph/0305196]}}].

\bibitem{icecube2}
{\scshape IceCube} collaboration, R.~Abbasi et~al., \emph{The design and
  performance of {IceCube DeepCore}},
  \href{http://dx.doi.org/10.1016/j.astropartphys.2012.01.004}{\emph{Astropart.
  Phys.} {\bf 35} (2012) 615--624},
  [\href{http://arxiv.org/abs/arXiv:1109.6096}{{[\tt arXiv:1109.6096]}}].

\bibitem{superk}
S.~Fukuda et~al., \emph{The super-kamiokande detector},
  \href{http://dx.doi.org/http://dx.doi.org/10.1016/S0168-9002(03)00425-X}{\emph{Nucl.
  Instrum. Meth. A} {\bf 501} (2003) 418 -- 462}.

\bibitem{Guo:2007ug}
{\scshape Daya Bay} collaboration, X.~Guo et~al., \emph{{A Precision
  measurement of the neutrino mixing angle $\theta_{13}$ using reactor
  antineutrinos at Daya-Bay}},
  \href{http://arxiv.org/abs/hep-ex/0701029}{{[\tt hep-ex/0701029]}}.

\bibitem{Racah:2016gnm}
E.~Racah, S.~Ko, P.~Sadowski, W.~Bhimji, C.~Tull, S.-Y. Oh et~al.,
  \emph{{Revealing Fundamental Physics from the Daya Bay Neutrino Experiment
  using Deep Neural Networks}},
  \href{http://arxiv.org/abs/arXiv:1601.07621}{{[\tt arXiv:1601.07621]}}.

\bibitem{Aad:2008zzm}
{\scshape ATLAS} collaboration, G.~Aad et~al., \emph{{The ATLAS Experiment at
  the CERN Large Hadron Collider}}, {\emph{JINST} {\bf 3} (2008) S08003}.

\bibitem{Chatrchyan:2008aa}
{\scshape CMS} collaboration, S.~Chatrchyan et~al., \emph{{The CMS experiment
  at the CERN LHC}}, {\emph{JINST} {\bf 3} (2008) S08004}.

\bibitem{deOliveira:2015xxd}
L.~de~Oliveira, M.~Kagan, L.~Mackey, B.~Nachman and A.~Schwartzman,
  \emph{{Jet-Images -- Deep Learning Edition}},
  \href{http://arxiv.org/abs/arXiv:1511.05190}{{[\tt arXiv:1511.05190]}}.

\bibitem{Hornik}
K.~Hornik, M.~Stinchcombe and H.~White, \emph{Multilayer feedforward networks
  are universal approximators},
  \href{http://dx.doi.org/http://dx.doi.org/10.1016/0893-6080(89)90020-8}{\emph{Neural
  Networks} {\bf 2} (1989) 359 -- 366}.

\bibitem{Cybenko}
G.~Cybenko, \emph{Approximation by superposition of a sigmoidal function},
  {\emph{Math. Control Signals System} {\bf 2} (1989) 303--314}.

\bibitem{Rumelhart:1986:PDP:104279}
D.~E. Rumelhart, J.~L. McClelland and C.~PDP Research~Group, eds.,
  \emph{Parallel Distributed Processing: Explorations in the Microstructure of
  Cognition, Vol. 1: Foundations}.
\newblock MIT Press, 1986.

\bibitem{hinton1986}
D.~E. Rumelhart, G.~E. Hinton and R.~J. Williams, \emph{{Learning
  representations by back-propagating errors}}, {\emph{Nature} {\bf 323} (1986)
  533--536+}.

\bibitem{LeCun1998}
Y.~LeCun, L.~Bottou, G.~B. Orr and K.~R. M{\"u}ller, \emph{Neural Networks:
  Tricks of the Trade}.
\newblock Springer Berlin Heidelberg, 1998.

\bibitem{Bengio-2009}
Y.~Bengio, \emph{Learning deep architectures for {AI}}, {\emph{Foundations and
  Trends in Machine Learning} {\bf 2} (2009) 1--127}.

\bibitem{lecun2015deep}
Y.~LeCun, Y.~Bengio and G.~Hinton, \emph{Deep learning},
  \href{http://dx.doi.org/10.1038/nature14539}{\emph{Nature} {\bf 521} (2015)
  436--444}.

\bibitem{krizhevsky2012imagenet}
A.~Krizhevsky, I.~Sutskever and G.~E. Hinton, \emph{Imagenet classification
  with deep convolutional neural networks},  in \emph{Advances in Neural
  Information Processing Systems 25}, pp.~1097--1105, 2012.

\bibitem{szegedy2014googlenet}
C.~Szegedy, W.~Liu, Y.~Jia, P.~Sermanet, S.~Reed, D.~Anguelov et~al.,
  \emph{Going deeper with convolutions},
  \href{http://arxiv.org/abs/arXiv:1409.4842}{{[\tt arXiv:1409.4842]}}.

\bibitem{farabet-pami-13}
C.~Farabet, C.~Couprie, L.~Najman and Y.~LeCun, \emph{Learning hierarchical
  features for scene labeling}, {\emph{IEEE Transactions on Pattern Analysis
  and Machine Intelligence} (August, 2013) }.

\bibitem{icml2010_NairH10}
V.~Nair and G.~E. Hinton, \emph{Rectified linear units improve restricted
  boltzmann machines},  in \emph{Proceedings of the 27th International
  Conference on Machine Learning ({ICML-10})}, pp.~807--814, 2010.

\bibitem{hinton2014dropout}
N.~Srivastava, G.~Hinton, A.~Krizhevsky, I.~Sutskever and R.~Salakhutdinov,
  \emph{Dropout: A simple way to prevent neural networks from overfitting},
  {\emph{Journal of Machine Learning Research} {\bf 15} (2014) 1929--1958}.

\bibitem{lecun2010convolutional}
Y.~LeCun, K.~Kavukcuoglu and C.~Farabet, \emph{Convolutional networks and
  applications in vision},  in \emph{Circuits and Systems ({ISCAS}),
  Proceedings of 2010 {IEEE} International Symposium on}, pp.~253--256, 2010.
\newblock \href{http://dx.doi.org/10.1109/ISCAS.2010.5537907}{DOI}.

\bibitem{hubel68}
D.~Hubel and T.~Wiesel, \emph{Receptive fields and functional architecture of
  monkey striate cortex}, {\emph{Journal of Physiology} {\bf 195} (1968)
  215--243}.

\bibitem{ILSVRC15}
O.~Russakovsky, J.~Deng, H.~Su, J.~Krause, S.~Satheesh, S.~Ma et~al.,
  \emph{{ImageNet Large Scale Visual Recognition Challenge}},
  \href{http://dx.doi.org/10.1007/s11263-015-0816-y}{\emph{International
  Journal of Computer Vision (IJCV)} {\bf 115} (2015) 211--252}.

\bibitem{lin2013network}
M.~Lin, Q.~Chen and S.~Yan, \emph{Network in network},  in \emph{International
  Conference on Learning Representations}, 2014.
\newblock \href{http://arxiv.org/abs/arXiv:1312.4400}{{[\tt arXiv:1312.4400]}}.

\bibitem{numi}
K.~Anderson et~al., \emph{{The NuMI Facility Technical Design Report}},
  FERMILAB-DESIGN-1998-01.

\bibitem{caffe}
Y.~Jia, E.~Shelhamer, J.~Donahue, S.~Karayev, J.~Long, R.~Girshick et~al.,
  \emph{Caffe: Convolutional architecture for fast feature embedding},
  \href{http://arxiv.org/abs/arXiv:1408.5093}{{[\tt arXiv:1408.5093]}}.

\bibitem{Bishop:2006:PRM:1162264}
C.~M. Bishop, \emph{Pattern Recognition and Machine Learning (Information
  Science and Statistics)}.
\newblock Springer-Verlag New York, Inc., 2006.

\bibitem{fluka1}
A.~Ferrari, P.~R. Sala, A.~Fass{\`o} and J.~Ranft, \emph{{FLUKA}: A
  multi-particle transport code (program version 2005)},  CERN-2005-010,
  SLAC-R-773, INFN-TC-05-11.

\bibitem{fluka2}
T.~T. B{\"o}hlen, F.~Cerutti, M.~P.~W. Chin, A.~Fass{\`o}, A.~Ferrari, P.~G.
  Ortega et~al., \emph{The {FLUKA} code: Developments and challenges for high
  energy and medical applications},
  \href{http://dx.doi.org/http://dx.doi.org/10.1016/j.nds.2014.07.049}{\emph{Nuclear
  Data Sheets} {\bf 120} (2014) 211 -- 214}.

\bibitem{beamline}
P.~Adamson et~al., \emph{{The NuMI Neutrino Beam}},
  \href{http://dx.doi.org/10.1016/j.nima.2015.08.063}{\emph{Nucl. Instrum.
  Meth.} {\bf A806} (2016) 279--306},
  [\href{http://arxiv.org/abs/1507.06690}{{[\tt 1507.06690]}}].

\bibitem{genie}
C.~Andreopoulos et~al., \emph{The {GENIE} neutrino monte carlo generator},
  {\emph{Nucl. Instrum. Meth.} {\bf A614} (2010) 87--104}.

\bibitem{geant1}
J.~Allison et~al., \emph{Geant4 developments and applications}, {\emph{IEEE
  Trans. Nucl. Sci.} {\bf 53} (2006) 270--278}.

\bibitem{geant2}
S.~Agostinelli et~al., \emph{Geant4 - a simulation toolkit}, {\emph{Nucl.
  Instrum. Meth.} {\bf A506} (2003) 250--303}.

\bibitem{novaSim}
A.~Aurisano, C.~Backhouse, R.~Hatcher, N.~Mayer, J.~Musser, R.~Patterson
  et~al., \emph{The {NOvA} simulation chain},
  \href{http://dx.doi.org/10.1088/1742-6596/664/7/072002}{\emph{J. Phys. Conf.
  Ser.} {\bf 664} (2015) 072002} FERMILAB-CONF-15-203-ND.

\bibitem{GVK218647751}
J.~Long, \emph{Regression models for categorical and limited dependent
  variables}.
\newblock Sage Publ., 1997.

\bibitem{cuda}
J.~Nickolls, I.~Buck, M.~Garland and K.~Skadron, \emph{Scalable parallel
  programming with cuda},  in \emph{Que}, 2008.

\bibitem{pdg}
{\scshape Particle Data Group} collaboration, K.~A. Olive et~al., \emph{Review
  of particle physics},
  \href{http://dx.doi.org/10.1088/1674-1137/38/9/090001}{\emph{Chin. Phys.}
  {\bf C38} (2014) 090001}.

\bibitem{crust}
C.~Bassin, G.~Laske and G.~Masters, \emph{The current limits of resolution for
  surface wave tomography in north america}, {\emph{EOS Trans. AGU} } 81, F897,
  2000.

\bibitem{Adamson:2016tbq}
P.~Adamson et~al., \emph{First measurement of electron neutrino appearance in
  {NOvA}},  \href{http://arxiv.org/abs/arXiv:1601.05022}{{[\tt
  arXiv:1601.05022]}}.

\bibitem{Adamson:2016xxw}
P.~Adamson et~al., \emph{First measurement of muon-neutrino disappearance in
  {NOvA}}, {\emph{Phys. Rev. D} {\bf 93} (2016) 051104}.

\bibitem{Baird:2015pgm}
M.~Baird, J.~Bian, M.~Messier, E.~Niner, D.~Rocco and K.~Sachdev, \emph{{Event
  Reconstruction Techniques in NOvA}},
  \href{http://dx.doi.org/10.1088/1742-6596/664/7/072035}{\emph{J. Phys. Conf.
  Ser.} {\bf 664} (2015) 072035}.

\bibitem{DBLP:journals/corr/LongSD14}
J.~Long, E.~Shelhamer and T.~Darrell, \emph{Fully convolutional networks for
  semantic segmentation},  in \emph{Conference on Computer Vision and Pattern
  Recognition}, 2015.
\newblock \href{http://arxiv.org/abs/arXiv:1411.4038}{{[\tt arXiv:1411.4038]}}.

\end{thebibliography}\endgroup

\end{document}